\documentclass[fleqn,usenatbib]{mnras}

\usepackage{ae,aecompl}

\usepackage{graphicx}	
\usepackage{amsmath}	
\usepackage{amssymb}	

\usepackage{color}
\usepackage{subfigure}
\usepackage{multirow}
\usepackage{float}

\def\sec{Section~}
\def\fig{Fig.~} 
  
\def\eq{equation~} 
\def\tab{Table~}
\defcitealias{Recchia-2016-08082016}{Paper I}

\title[]{Cosmic ray driven winds in the Galactic environment and the cosmic ray spectrum}

\author[]{
	S. Recchia $^{1}$\thanks{E-mail: sarah.recchia@gssi.infn.it}
	P. Blasi,$^{2,1}$\thanks{E-mail: blasi@arcetri.astro.it}
	G. Morlino $^{1,2}$\thanks{E-mail: giovanni.morlino@gssi.infn.it}
	\\
	$^{1}$ Gran Sasso Science Institute (INFN), Viale F. Crispi 7 - 67100 L' Aquila, Italy\\
	$^{2}$ INAF/Osservatorio Astrofisico di Arcetri, Largo E. Fermi, 5 - 50125 Firenze, Italy
}

\date{Accepted XXX. Received YYY; in original form ZZZ}

\pubyear{2016}

\begin{document}

\label{firstpage}
\pagerange{\pageref{firstpage}--\pageref{lastpage}}
\maketitle

\begin{abstract}
Cosmic Rays escaping the Galaxy exert a force on the interstellar medium directed away from the Galactic disk. If this force is larger than the gravitational pull due to the mass embedded in the Galaxy, then galactic winds may be launched. Such outflows may have important implications for the history of star formation of the host galaxy, and in turn affect in a crucial way the transport of cosmic rays, both due to advection with the wind and to the excitation of waves by the same cosmic rays, through streaming instability. The possibility to launch cosmic ray induced winds and the properties of such winds depend on environmental conditions, such as the density and temperature of the plasma at the base of the wind and the gravitational potential, especially the one contributed by the dark matter halo. In this paper we make a critical assessment of the possibility to launch cosmic ray induced winds for a Milky-Way-like galaxy and how the properties of the wind depend upon the conditions at the base of the wind. Special attention is devoted to the implications of different conditions for wind launching on the spectrum of cosmic rays observed at different locations in the disc of the galaxy. We also comment on how cosmic ray induced winds compare with recent observations of Oxygen absorption lines in quasar spectra and emission lines from blank-sky, as measured by XMM-Newton/EPIC-MOS.  
\end{abstract}

\begin{keywords}
keyword1 -- keyword2 -- keyword3
\end{keywords}



\section{Introduction}
Galactic winds have been observed in many galaxies (see e.g  \cite{Veilleux-2005ARA&A..43..769V}; \cite{Martin-2012}; \cite{King-Pound-2015ARA&A..53..115K}) and constitute an important feedback process to the evolution of galaxies. In fact, by removing gas from the galactic disk galactic winds regulate the star formation rate (see e.g. \cite{Crain-2007MNRAS.377...41C}) and pollute  galactic halos and  intergalactic space with hot plasma and metals, thus affecting the chemical composition, the temperature and the degree of ionization of the interstellar medium (ISM) and of the intergalactic medium (IGM) (see e.g \cite{Miller-2015-0004-637X-800-1-14}). Moreover, the expelled material may contain an appreciable fraction of the number of baryons in the Universe (see e.g. \cite{Kalberla-2008A&A...487..951K}; \cite{Miller-2015-0004-637X-800-1-14}) and may contribute to solving the problem of missing baryons in the local Universe (see e.g \cite{Michael-2010}).
Although observations of the Milky Way have not yet led to conclusive evidence of the existence of a Galactic wind, hot dilute gas, possibly connected with such winds, has been detected through the continuous emission in the X-ray band (\cite{Breitschwerdt-1994Natur.371..774B}; \cite{Breitschwerdt-1999A&A...347..650B}; \cite{Everett-2008p3580}) and through Oxygen absorption lines ($O_{VII}$ and $O_{VIII}$ lines) in the  spectrum of distant Quasars (\cite{Miller-2015-0004-637X-800-1-14}).
The features of such lines are compatible with a halo with a mass of $\sim 10^{10} \rm M_{\odot}$, an extension of  $\sim 100$ kpc, a temperature in the range $10^6-10^7$ K and an average metallicity of Z $\sim 0.2-0.3$.
In particular the latter data imply that the gas in the halo comes from the Galaxy and not from intergalactic space (\cite{Miller-2015-0004-637X-800-1-14}). Moreover, the recent observation of the so called Fermi Bubbles may be associated to recent bursts near the Galactic Center region (see \cite{Cheng-2011ApJ...731L..17C}; \cite{Zubovas-2011MNRAS.415L..21Z}) or to past starburst activities (see \cite{Lacki-2014MNRAS.444L..39L}).

Galactic winds may be 1) thermally-driven, namely generated  by the heating of the ISM due to SN explosions (see e.g \cite{Chevalier-1985Natur.317...44C}) or by accretion onto the super-massive black holes in the center of AGN (see e.g \cite{King-Pound-2015ARA&A..53..115K});  2) radiation pressure-driven (see e.g \cite{Scoville-2003JKAS...36..167S}; \cite{Murray-2005ApJ...618..569M}); 3) CR-driven, namely due to the pressure exerted by escaping CRs on the ISM (see e.g \cite{Ipavich-1975p3566} \cite{Breitschwerdt-1991A&A...245...79B}; \cite{Uhlig-2012MNRAS.423.2374U}). The first two  mechanisms are likely to take place in  starburst galaxies and galaxies with active nuclei.

In the Milky Way the thermal and radiation pressure gradients  are expected to be too small (except perhaps for the Galactic Center region) to drive outflows.  However, CRs  escaping from the Galaxy lead to a gradient in the CR density with several implications. First, the CR pressure gradient exerts a force, $-\nabla P_c$ , on the background plasma directed away from the disk and opposite to the Galactic gravitational force (due to both baryonic and dark matter). If this force is larger than the gravitational pull it may launch a wind. Second, the CR density gradient can excite Alfv\'{e}n waves,  due to CR streaming instability (\cite{Kulsrud-1971}; \cite{Skilling-1975}), which move in the direction of the decreasing CR density. Such waves affect the scattering properties of CRs, namely their diffusive transport. Third, winds and the waves that propagate inside the outgoing plasma, advect CRs and contribute to change the spectrum of CRs as observed in the Galaxy.

CRs have been widely recognized as an appealing way to launch winds in our Galaxy  and the hydrodynamics of CR-driven winds has been extensively studied in stationary one dimensional calculations (see e.g \cite{Ipavich-1975p3566}; \cite{Breitschwerdt-1991A&A...245...79B}; \cite{Everett-2008p3580}; \cite{Zirakashvili-1996A&A...311..113Z}), in time-dependent calculations (\cite{Dorfi-2012A&A...540A..77D}), in hydrodynamical simulations (see \cite{Uhlig-2012MNRAS.423.2374U}; \cite{Booth-2013ApJ...777L..16B}; \cite{Salem-2014MNRAS.437.3312S}) and in MHD simulations (see \cite{Girichidis-2016ApJ...816L..19G}; \cite{Ruszkowski-2016}). The paper by \cite{Breitschwerdt-1991A&A...245...79B} represented a crucial step forward in the investigation of CR-driven Galactic winds: their computation techniques have been adopted by much of the subsequent work on the subject. 

In all this bulk of work, no matter the level of refinement, only the hydrodynamics of CR-driven winds was addressed, while no information was retained on the CR spectrum in such winds. On the other hand, the CR spectrum represents a crucial  observable that we can use in order to constraint wind models, hence its prediction is essential. Recently, our group \cite[hereafter Paper I]{Recchia-2016-08082016}  presented an extensive study of the problem of the CR transport in a CR-driven wind. We developed a semi-analytical method which allows to compute self-consistently the hydrodynamical properties of CR-driven winds and the corresponding CR distribution function in the presence of self-generated Alfv\'{e}n waves. In our approach, the hydrodynamics of the wind is calculated following the one-dimensional stationary model developed by \cite{Breitschwerdt-1991A&A...245...79B} and the CR transport is due to scattering of CRs on the self-generated turbulence and to the CR advection with such waves and with the wind. Both the flow and the CR transport only occur along the magnetic field lines and the flow geometry is preassigned.
The CR diffusion coefficient is calculated from the spectrum of waves excited by streaming instability and locally damped through non linear Landau damping (NLLD). The importance of CRs for  wind launching was confirmed by our work, although we found that in general, the CR spectrum in the presence of winds launched in the vicinity of the Sun is not in agreement with observations. This is mainly due to the fact that CR advection in such winds is strong, leading to spectra at low energies (below $\sim 200$ GeV) which are harder than the observed spectrum. At high energy (above $\sim 200$ GeV), instead, the wind expansion together with the steep energy dependence of the self-generated diffusion coefficient, leads to spectra that are steeper than the observed spectrum. However, these conclusions strongly depend, at low energy, on the wind launching parameters and, at high energy, on the physical conditions in the near-disk region, in particular on the amount of neutral gas in such region and on its volume filling factor. 
For instance, we showed that at high energy the CR spectrum is very sensitive to the diffusion properties of the near-disk region ($\lesssim$ 1 kpc) where the presence of a pre-existing turbulence would lead to a spectral hardening  in the high energy part of the spectrum (see also \cite{Aloisio-2015p3650}), in agreement with  recent observations of the proton and helium spectrum (see \cite{pamhard}; \cite{amshard}).

In this paper we use the calculations developed in \citetalias{Recchia-2016-08082016}   and we study the possibility to launch winds, the dependence of the wind properties on environmental parameters and the implications of such dependences for the spectrum of CRs observed at the Sun position. Special attention is devoted to investigating the wind properties as functions of the ISM properties (gas density and temperature, Galactic magnetic field), of the CR pressure, of the flow geometry and on the Galactic gravitational potential.

The spatial dependence of these quantities in the Galaxy is constrained by observations (see e.g \cite{Ferriere-2001}; \cite{Cox-2005}; \cite{Wolfire-2003ApJ...587..278W}; \cite{Miller-2015-0004-637X-800-1-14}). In particular,  the observation of the Oxygen absorption lines in the  spectrum of distant Quasars analyzed by \cite{Miller-2015-0004-637X-800-1-14} suggests the presence of a large Galactic halo, presumably due to a Galactic wind, and provides us with indications on the gas mass distribution and temperature of such halo, and consequently on the flow geometry. Moreover, the baryonic and DM content of the Milky Way and the corresponding gravitational potential can be constrained by analyzing the dynamics of stars and gas (see e.g \cite{Bertone-2005PhR...405..279B}; \cite{Sofue-2012PASJ...64...75S}; \cite{Salucci-2013JCAP...07..016N}). We discuss how our results change by adopting different distributions of Galactic dark matter and compare them with the halo structure deduced by \cite{Miller-2015-0004-637X-800-1-14}.\\

The paper is organized as follows: in \sec \ref{sec:wind-model} we describe our wind model and we briefly summarize the method used to solve the hydrodynamic wind equations and the CR transport equation, as illustrated in \citetalias{Recchia-2016-08082016}. In \sec \ref{sec:grav-pot} we describe the model used for the gravitational potential of the Galaxy, including several  DM halo models. In \sec \ref{sec:input-param} we summarize the main  properties of the ISM as deduced from observations and their relevance for the models of CR driven winds. In \sec \ref{sec:param-study}, we discuss in detail the dependence of the wind properties on the parameters of the problem, as well as on the geometry of the outflow and the Galactic gravitational potential. Finally, in \sec \ref{sec:spectrum}, we illustrate the implications of CR driven winds launched at the position of the Sun on the spectrum of CRs. 

\section{The wind model}
\label{sec:wind-model}
The dynamics of a CR-driven wind in the stationary regime is described by the system of hydrodynamic equations for the wind (i.e the equations of mass, momentum and energy conservation) and by the kinetic transport equation for CRs. The hydrodynamic wind equations have been written down and discussed in detail by \cite{Breitschwerdt-1991A&A...245...79B}. A method to calculate the combined solution of the hydrodynamical equations for the wind and the transport equation of CRs has been recently developed in \citetalias{Recchia-2016-08082016}. This method allows one to determine at the same time the wind structure and the related CR spatial and momentum distribution. Here we briefly summarize the equations and the main features of the wind model.

Following \cite{Breitschwerdt-1991A&A...245...79B}, we assume that the wind is launched from a surface at distance $z_{0}$ above (and below) the Galactic disc. The geometry of the outflow induced by the CR pressure gradient is assumed to be pre-assigned and reflecting the geometry of magnetic field lines. In our calculation the CR diffusion coefficient is due to scattering off self-generated Alfv\'{e}n waves, as due to the excitation of streaming instability, saturated through NLLD. Advection of CRs with self-generated waves and with the wind are taken into account. Wave damping is considered to occur on a timescale much shorter than any other considered process (definitely justified for NLLD) and contribute to the gas heating. Under these assumptions, as shown in \citetalias{Recchia-2016-08082016}, the hydrodynamic wind equations and the CR transport equation become effectively one-dimensional and only depend on the distance $z$ from the Galactic disk. 
The hydrodynamic equations read
\begin{align}      
& \rho u A=\textnormal{const}, \label{eq:Hcal-mass} \\
& AB= \textnormal{const},\label{eq:Hcal-B}\\
& \frac{du}{dz}=u\frac{c_{*}^2\frac{1}{A} \frac{dA}{dz} -\frac{d\Phi}{dz}}{u^2-c_{*}^2}, \label{eq:Hcal-wind eq}\\
& \frac{dP_g}{dz}=\gamma_g\frac{P_g}{\rho}\frac{d\rho}{dz}-(\gamma_g -1)\frac{v_A}{u}\frac{dP_c}{dz} \label{eq:Hcal-Pg}\\
&\frac{dP_c}{dz}= \gamma_{eff}\frac{P_c}{\rho}\frac{2u+v_A}{2(u+v_A)}\frac{d\rho}{dz}, \label{eq:Hcal-Pc}\\
& c_{*}^2=\gamma_g \frac{P_g}{\rho} + \gamma_{eff} \frac{P_c}{\rho} \left[ 1-(\gamma_g-1)\frac{v_A}{u} \right]\frac{2u+v_A}{2(u+v_A)}, \label{eq:Hcal-cs}\\
& \frac{\gamma_{eff}}{\gamma_{eff} -1}  =  \frac{\gamma_{c}}{\gamma_{c} -1} -\frac{\overline{D}}{(\gamma_c -1)(u+v_A)P_c}\frac{dP_c}{dz}. \label{eq:Hcal-GAeff}
\end{align}
where $\rho(z)$, $u(z)$ and $P_g(z)$ are the gas density, velocity and pressure,  $B(z)$ is the  background magnetic field strength and $\gamma_g = 5/3$ is the adiabatic index of the gas.
$P_{c}(z)$ is the CR pressure and $\gamma_{c}(z)$ is the adiabatic index of the CR gas, while the average  CR diffusion coefficient $\overline{D}$ is defined in \eq \ref{eq:Db-def}.  Notice that $\gamma_{c}(z)$ is actually calculated as a function of $z$ from the distribution function $f(p, z)$ that solves the CR transport equation. The space-dependent Alfv\'{e}n velocity is $v_A(z) = B(z)/\sqrt{4\pi\rho(z)}$, while $\Phi(z)$ is the gravitational potential of the Galaxy. The wave pressure does not appear explicitly since the assumption of fast damping results in a wave pressure much smaller than the gas and CR pressure.\\
Two important quantities appear in these equations: the generalized sound speed $c_{*}$ and the  effective adiabatic index for CRs, $\gamma_{eff}$. Looking at the  definition of $c_{*}^2$, \eq \ref{eq:Hcal-cs}, one can recognize the sum of the sound speed term (square root of $\gamma_g P_g/\rho$) and of a "CR sound speed" term (the term $\gamma_{eff} P_c/\rho$  is formally identical to the  square root of the sound speed) which depends on the Alfv\'{e}nic Mach number $u/v_A$. Notice that the term $-(\gamma_g -1)v_A/u$ is due to the non adiabatic gas heating induced by wave damping in \eq \ref{eq:Hcal-Pg}. The importance of $c_{*}$ resides in the fact that $u=c_{*}$ is the condition for the flow to become sonic but also is a critical point for the wind equation \ref{eq:Hcal-wind eq}.
The effective adiabatic index for CRs, $\gamma_{eff}$, takes into account the CR diffusivity in the hydrodynamic equations (see also \cite{Zirakashvili-1996A&A...311..113Z}). In general, $\gamma_{eff}$ is a function of $z$, however,  it has been verified  that in general (including in the cases presented here) $\gamma_{eff}$ shows a weak dependence on $z$ and we can safely treat it as a constant. Typical values of $\gamma_{eff}\sim  1.1-1.2 $, while $\gamma_c \sim 4/3$.
\cite{Breitschwerdt-1991A&A...245...79B} also introduced the generalized sound speed $c_{*}$ but with two important differences compared with the calculation discussed here. First, they neglected the CR diffusion by assuming  that $\overline{D}=0$ so that in their calculation  $\gamma_{c}$ appears instead of $\gamma_{eff}$. Neglecting diffusion is not a good approximation, at least not at all distances $z$ from the disc. Second, \cite{Breitschwerdt-1991A&A...245...79B} did not take wave damping into account and retained the wave pressure in the hydrodynamic equations. Without damping, the growth of waves due to CR streaming can in fact lead to a wave pressure which is comparable with the CR pressure. For this reason, instead of having the term $-(\gamma_g -1)v_A/u$ in the definition of $c_{*}^2$, they have a "wave sound speed" term which takes into account the effect of the wave pressure.

The CR transport equation reads
\begin{equation}\label{eq: CR transport equation flux tube}
\frac{\partial}{\partial z} \left[ A\,D\frac{\partial f}{\partial z}\right] - A\,U\frac{\partial f}{\partial z}  +\frac{d(A\,U)}{dz}\frac{1}{3}\frac{\partial f}{\partial \ln p}+A\,Q=0,
\end{equation}
where $f(z,p)$ and $D(z,p)$ are the CR distribution function and diffusion coefficient as functions of position $z$ and momentum $p$ and $U(z)=u(z)+v_{A}(z)$ is the advection velocity including the wind speed and the Alfv\'{e}n speed.

The term $Q(z,p)$ represents the injection of CRs in the Galaxy, which we assume to be limited to the Galactic disc (considered as infinitely thin) and can be written as  $Q(z,p)=Q_0(p)\delta(z)$. The function $Q_0(p)$ describes the injection spectrum,
\begin{equation}\label{eq:injection}
Q_0(p)=\frac{\mathcal{N}_{SN}(p)\mathcal{R}_{SN}}{\pi R_d^2},
\end{equation}
where $R_d$ is the Galactic disk  and $\mathcal{N}_{SN}(p)$ is the spectrum contributed by individual supernovae (SN) occurring at a rate $\mathcal{R}_{SN}$ and reads 
\begin{equation}\label{eq:SN-spectrum}
\mathcal{N}_{SN}(p)=\frac{\xi_{CR}E_{SN}}{I(\gamma)c(mc)^4}\left(\frac{p}{mc} \right)^{-\gamma}.
\end{equation}
In the last expression $\xi_{CR}$ is the CR injection efficiency (typically $\sim 10\%$), $E_{SN}$ is the energy released by a SN explosion ($\sim 10^{51}$erg), $\mathcal{R}_{SN}$ is the rate of SN explosions ($\sim 1/30 \, $yr$ ^{-1}$), and $I(\gamma)$ is a normalization factor (see \citetalias{Recchia-2016-08082016}).\\
The average diffusion coefficient of \eq \ref{eq:Hcal-GAeff} is defined as:
\begin{equation}\label{eq:Db-def}
\overline{D} (z)= \frac{\int_{0}^{\infty} dp~ p^{2} T(p) D(z,p) \nabla f}{\int_{0}^{\infty} dp ~p^{2} T(p) \nabla f},
\end{equation}
where $T(p)$ is the kinetic energy of particles with momentum $p$. The diffusion coefficient $D(z,p)$ is determined by the local balance between the CR-driven growth of Alfv\'{e}n waves and their damping through NLLD. Since the self-generated perturbations in the magnetic field are relatively weak, one can use quasi-linear theory to write the diffusion coefficient as:
\begin{equation}\label{eq:D self-gen}
D(z,p)=\left.\frac{1}{3}\frac{v(p)r_L(z,p)}{\mathcal{F}(z, k_{res})}\right|_{k_{\rm res}=1/r_L},
\end{equation}
where $\mathcal{F}$ is the normalized energy density per unit logarithmic wavenumber $k$, calculated at the resonant wavenumber $k_{\rm res}=1/r_L(p)$. In the regions where the background gas is totally ionized, waves are damped through NLLD at a rate (see \cite{Zhou-nlld}; \cite{Ptuskin-2003A&A...403....1P}):
\begin{equation}
\Gamma_{\rm D} = (2c_k)^{-3/2}kv_A \mathcal{F}^{1/2}, \label{eq:GAMMA NLD} 
\end{equation}
where $c_k = 3.6$.
On the other hand the CR-driven growth occurs at a rate which is given by \cite{Skilling-1971p2173}:
\begin{equation}
\Gamma_{\rm CR}=\frac{16\pi^2}{3}\frac{v_A}{\mathcal{F} B^2}\left[p^4v(p)\left| \frac{\partial f}{\partial z} \right| \right]_{p=p_{res}}. \label{eq:GAMMA CR}
\end{equation}
Equating the two rates, $\Gamma_{\rm D}= \Gamma_{\rm CR}$, one obtains:
\begin{equation}\label{eq:wave spectrum}
\mathcal{F}(z,p)=2c_k\left[\frac{p^4v(p) \left| \frac{\partial f}{\partial z} \right| 
	\frac{16\pi^2}{3}r_L(z,p)}{B^2(z)} \right]^{2/3},
\end{equation}
where $\mathcal{F}$ is expressed as a function of momentum by means of the resonant condition $p = p_{res}(k)$.\\

As for the flow geometry, we adopt one that is similar to the one introduced by \cite{Breitschwerdt-1991A&A...245...79B} namely we assume that the wind is launched at the wind base $z_0$ and that it retains a cylindrical geometry out to a characteristic distance $Z_{b}$.  At larger distances the flow opens up as $\sim\,z^{\alpha}$. The wind  area transverse to $z$ is then assumed to be in the form:
\begin{equation}\label{eq:area}
A(z)=A_{0} \left[ 1+ \left( \frac{z}{Z_b}\right)^{\alpha}\right],
\end{equation}
which is only function of the spatial coordinate $z$, namely of the distance from the Galactic disk. Notice that in \cite{Breitschwerdt-1991A&A...245...79B} and in \citetalias{Recchia-2016-08082016} the authors assumed $Z_b \sim 15$ kpc and $\alpha = 2$, which corresponds to spherical opening. Here, instead, we explore the implications of changing these two parameters for the wind properties.

Following  \citetalias{Recchia-2016-08082016}, the hydrodynamic equations for the wind and the CR transport equation are solved together following at iteration scheme. 
For a given set  of input parameters at the wind base $z_0$ (namely the gas density $\rho_0$, the gas temperature $T_0$, the magnetic field strength $B_0$ and the CR pressure $P_{c0}$), for each iteration we determine the velocity $u_0$ at the base of the wind, for which the flow experiences a smooth transition from the subsonic to the supersonic regime. Given $u_0$, both the mass and energy flux of the wind are fixed, and it is possible to compute all the hydrodynamic quantities (gas density and pressure, wind velocity and CR pressure) as functions of the distance from the Galactic disk. For each iteration, the transport \eq \ref{eq: CR transport equation flux tube} is solved for the CR distribution function $f(z,p)$ and the diffusion coefficient $D(p,z)$. The iteration is then repeated until convergence is reached within a  given accuracy. 

\section{The Galactic gravitational potential}
\label{sec:grav-pot}
The mass distribution of the Milky Way, and of other galaxies as well,  is mainly inferred from analyses of the dynamics of stars and gas, namely by studying kinematic data such as rotation velocities, velocity dispersions and motion of satellite galaxies. Current models for the mass distribution of the Galaxy include a central bulge, a stellar and gas disk and a DM halo (see e.g \cite{Irrgang-2013A&A...549A.137I}).

Following \cite{Breitschwerdt-1991A&A...245...79B} and \cite{Irrgang-2013A&A...549A.137I}, the gravitational potential of the Galactic bulge and disk can be modelled using the parametrization of \cite{Miyamoto-1975PASJ...27..533M}: 
\begin{equation}\label{eq:bulge-disk}
\Phi_{\rm B,D}(R_0, z)= - \sum_{i=1}^2 \frac{G M_i}{\sqrt{R_0^2 + \left(a_i + \sqrt{z^2 + b_i^2}\right)^2}},
\end{equation}
where $z$ and $R_0$ are the distance from the Galactic disk and the Galactocentric distance respectively. 
For the parameters of this model we used the values proposed by \cite{Sofue-2012PASJ...64...75S}, namely 
$(M_1, a_1, b_1)=$($1.652\times 10^{10}M_{\odot}$, 0.0 kpc, 0.522 kpc) for the bulge and $(M_2, a_2, b_2)=$($3.4\times 10^{10}M_{\odot}$, 3.19 kpc, 0.289 kpc) for the disk.

For the dark matter mass distribution we considered three models, namely the  Navarro-Frenk-White \cite[]{nfw}, the Burkert  \cite[]{Burkert-1995ApJ...447L..25B} and the Einasto \cite[]{Retana-Montenegro-2012} profiles:

\begin{itemize}
	
	\item Navarro-Frenk-White (NFW) (\cite{nfw}):\\
	\\
	the density profile is of the form
	\begin{equation}
	\rho_{\rm NFW} = \frac{\rho_{0}}{x(1+x)^{2}},
	\end{equation}
	where $x=r/r_{c}$ and $r_{c}$ is the scale radius of the distribution. For the two quantities $\rho_{0}$ and $r_{c}$  we considered both the values given for the Milky Way by \cite{Sofue-2012PASJ...64...75S} ($\rho_{0}= 1.06 \times 10^7$ M$_{\odot}$ kpc$^{-3}$; $r_c = 12.0$ kpc) and by \cite{Salucci-2013JCAP...07..016N} ($\rho_{0}= 1.3 \times 10^7$ M$_{\odot}$ kpc$^{-3}$; $r_c = 16.0$ kpc). We refer to the two sets of parameters as NFW-Sofue and NFW-Salucci respectively. The DM halo is assumed to extend out to a maximum distance that equals the virial radius, which corresponds to $r_{\rm vir}\approx 240$ kpc for NFW-Sofue and  $r_{\rm vir}\approx 320$ kpc for NFW-Salucci.\\
	
	\item Burkert (BUR) (\cite{Burkert-1995ApJ...447L..25B}):\\ 
	\\
	the density profile is of the form
	\begin{equation}
	\rho_{\rm BUR} = \frac{\rho_{0}}{(1+x)(1+x^2)},
	\end{equation}
	where $x=r/r_{c}$ and $r_{c}$ is the core radius. For the two quantities $\rho_{0}$ and $r_{c}$ we considered the values given for the Milky Way by \cite{Salucci-2013JCAP...07..016N} ($\rho_{0}= 4.13 \times 10^7$ M$_{\odot}$ kpc$^{-3}$; $r_c = 9.3$ kpc). The virial radius for this model is $r_{\rm vir}\approx 300$ kpc.\\
	
	\item Einasto (EIN) (\cite{Retana-Montenegro-2012}):\\
	\\
	the density profile is of the form
	\begin{equation}
	\rho_{\rm EIN} = \rho_0 \exp(-x^{\alpha}),
	\end{equation}
	where $x=r/r_{c}$. For the three quantities $\rho_{0}$, $r_{c}$ and $\alpha$ we considered the values given for the Milky Way by \cite{Bernal-2012} ($\rho_{0}= 3.5 \times 10^{11}$ M$_{\odot}$ kpc$^{-3}$; $r_c = 6.7\times 10^{-6}$ kpc, $\alpha=$0.17).
\end{itemize}
These halo models predict a DM density at the Sun position in the range $0.2-0.4$ GeV/cm$^3$, in agreement with most of the current literature  (see \cite{Salucci-2010A&A...523A..83S} for a discussion on the local DM density).

\begin{figure}
	\includegraphics[width=\columnwidth]{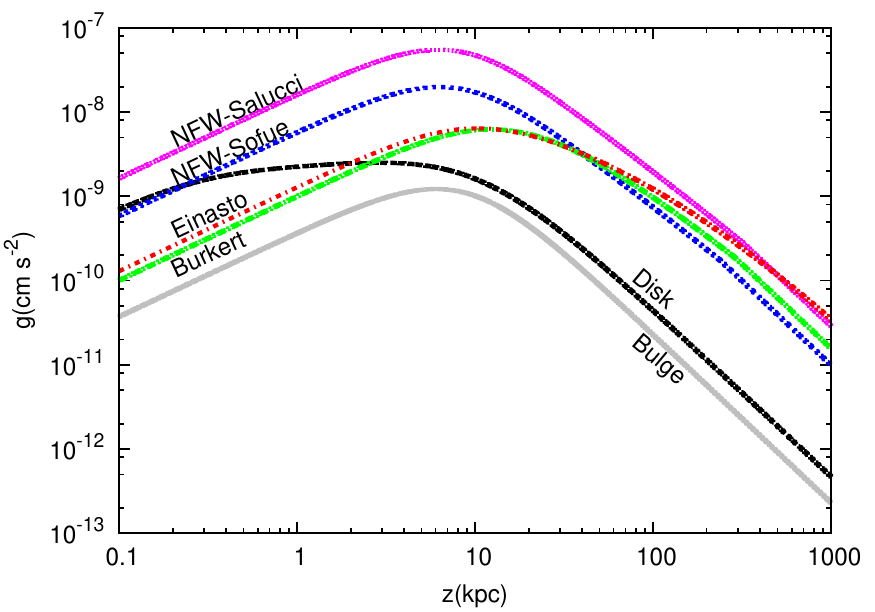}
	\caption{Galactic gravitational acceleration as function of the distance from the galactic plane, $z$, at the Sun position produced by different DM profiles. The contribution due to the disk and the bulge are also shown.}
	\label{fig:GravPot}
\end{figure}

A comparison between the gravitational acceleration produced by different DM models at the  solar galactocentric distance $R=R_{\odot}$ and as a function of the Galactic altitude  is shown in \fig \ref{fig:GravPot} where also the contribution due to the disk and the bulge are reported.
One can see that the Burkert and the Einasto profiles are quite similar in magnitude below  $z \sim 50$ kpc and that the Einasto profile becomes systematically larger than the Burkert profile above  $z \sim 50$ kpc. On the other hand, the NFW profile  shows large differences in magnitude, compared to the other two profiles,  below $z \sim 30-40$ kpc. Since these profiles predict a similar DM density at the Sun, this  is due to the different functional dependence on $x=r/r_{c}$ and to the different values of the scale radius $r_c$. 
\section{Fiducial values for the environmental parameters}
\label{sec:input-param}
The ISM is a rich environment in which several components, distinguished by the level of ionization and temperature, coexist and interact: molecular and atomic neutral gas, warm and hot ionized gas, interstellar magnetic field and CRs. 
In the cooler, denser regions of the ISM, matter is primarily in molecular form and can reach densities as high as $10^6$ cm$^{-3}$. On the other hand, in the diffuse regions of the ISM, matter is mainly ionized and can reach densities  as low as $10^{-4}$ cm$^{-3}$. In addition, the colder and denser phases of the ISM are mainly confined in the Galactic disk and have filling factor of order $\sim 1-5 \%$, much smaller than the dilute hot gas filling factor, which is of order $30-70 \%$. A detailed study of the gas components of the ISM has been carried out by several authors (see e.g \cite{Cox-2005}; \cite{Wolfire-2003ApJ...587..278W}) while a comprehensive review of the present knowledge of the ISM can be found in (\cite{Ferriere-2001}). In \tab \ref{tab:ISM-comp} we summarize the main properties of the gas components of the ISM in the vicinity of the Sun as reported by \cite{Ferriere-2001}. \\
\begin{table}
\caption{Components of the ISM and their properties in the vicinity of the Sun (see \protect\cite{Ferriere-2001}).} 
\label{tab:ISM-comp} 
\begin{center}
\begin{tabular}{|c|c|c|c|}
		\hline
		& &  & \\ 
		Component	& Temperature & Density & Scale Height\\
		& (K) &  (cm$^{-3}$)  & (kpc) \\
		& &  & \\ 
		\hline
		Molecular	& 10-20 & $10^2-10^6$ & $\sim 0.1$  \\ 
		\hline
		Cold Atomic	& 50-100 & 20-50 & $\sim 0.1-0.4$ \\ 
		\hline
		Warm Atomic	& 6000-10000 & 0.2-0.5 & $\sim 0.1-0.4$ \\ 
		\hline
		Warm Ionized	& 8000 & 0.2-0.5 & $\sim$ 1 \\ 
		\hline
		Hot Ionized	& $\sim 10^6$  & $\sim$ 0.006 & $> 1$ \\
		\hline
	\end{tabular}
\end{center}
\end{table}
The Galactic magnetic field shows a rich structure in which at least a disk component, which follows the spiral arms, and an out of disk component can be found (see \cite{Jansson-2012}).  The regular field strength changes with the Galactocentric distance. An estimation of the magnetic field strength in the vicinity of the Sun is reported for instance in \cite{Jansson-2012}, which quote the value  $\sim 1-2 \mu$G for the disk component and $\sim 1\mu$G for the out of disk component. \cite{Ferriere-2001} quote a value $B_{\odot}\sim 1.5\mu$G, while \cite{Cox-2005} suggest a value of $\sim 3-5\,\mu$G (which includes all field components). Here we are mainly interested in the out of disk component. 

Recently \cite{Miller-2015-0004-637X-800-1-14} analyzed the Oxygen absorption lines in quasar spectra and emission lines from blank-sky regions, measured by XMM-Newton/EPIC-MOS, and inferred the presence of a hot gaseous halo which could in fact be the result of a Galactic wind. Constraints on the structure of the hot halo can be found by fitting a radial model,  
\begin{equation}
n(r) \approx n_0\left(\frac{r_c}{r}\right)^{3\beta},
\end{equation}
for the halo density distribution, from which the expected emission, to compare with observations, is computed. Here $r=\sqrt{R^2 + z^2}$, where $R$ is the Galactocentric distance and $z$ the distance from the Galactic disk. The best fit parameters for the halo density given by \cite{Miller-2015-0004-637X-800-1-14} are $\beta = 0.5 \pm 0.3$ and $n_0 r_c^{3\beta} = 1.35 \pm 0.24$ cm$^{-3}$kpc$^{3\beta}$. The authors also infer a nearly constant halo temperature of $\sim 2\times 10^6$ K and a sub-solar gas metallicity that decreases with $r$, but that also must be $\gtrsim 0.3 Z_{\odot}$ to be consistent with the pulsar dispersion measure toward the Large Magellanic Cloud.

This halo model implies a gas density of $ 3-6 \times 10^{-3}$ cm$^{-3}$ in the vicinity of the Sun, a density scaling with $z$ that reads  $n(z) \sim z^{-1.5}$, and a halo mass of $M(<50 \rm kpc) = (3.8 \pm 0.3) \times 10^9 \, M_{\odot}$ and $M(<250 \rm kpc) = (4.3 \pm 0.8) \times 10^{10} \, M_{\odot}$. Notice that such halo mass would account for $\lesssim 50\%$ of the Milky Way missing baryons.\\

All these pieces of information allow us to define reasonable fiducial values for the input parameters of our wind problem (see \tab \ref{tab:input-param} for a summary): we are interested in the hot dilute phase of the ISM, for which we adopt a density $3-6 \times 10^{-3}$ cm$^{-3}$ and a temperature $1-3\times 10^6$K, and in the out of disk regular magnetic field 
for which we retain a field strength $1-2 \mu$G.\\ 
The pressure in the form of CR protons measured at Earth is $\sim 4\times 10^{-13}$ erg/cm$^{-3}$.\\
Finally,  \cite{Breitschwerdt-1991A&A...245...79B}, \cite{Everett-2008p3580} and  \citetalias{Recchia-2016-08082016}  assumed that the flux-tube opens up spherically, namely $\alpha =2$ in \eq \ref{eq:area}, while the results of \cite{Miller-2015-0004-637X-800-1-14} suggest $\alpha \sim 1.5$. As for the area length scale $Z_b$, \cite{Breitschwerdt-1991A&A...245...79B} and \citetalias{Recchia-2016-08082016} adopted the value $15$ kpc, while \cite{Everett-2008p3580} treated $Z_b$ as a fitting parameter and allowed it to vary around $\sim 5$ kpc. Here we consider $\alpha \sim 1.5-2$ and $Z_b \sim 5-15$ kpc.
\begin{table}
\caption {Fiducial values for the wind input parameters at the Sun position.} \label{tab:input-param} 
\begin{center}
	\begin{tabular}{|c|c|}
		\hline
		& \\
		parameter	& fiducial range  \\
		& \\ 
		\hline
		area-$\alpha$	& 1.5-2.0 \\ 
		\hline
		area-$Z_b$	& 5-15 kpc  \\ 
		\hline
		gas density	& $3-6 \times 10^{-3}$ cm$^{-3}$ \\ 
		\hline
		gas temperature	& $1-3\times 10^6$K \\ 
		\hline
		regular B	&  $1-2 \mu$G\\
		\hline
		CR pressure & $4\times 10^{-13}$ erg/cm$^{-3}$\\
		\hline
	\end{tabular}
\end{center}
\end{table}

\section{Dependence of the wind properties on the Galactic environment}
\label{sec:param-study}
In this section we present a purely hydrodynamical analysis  of Galactic winds, launched at the Sun position, focusing on how the  wind's properties are affected when changing the environmental parameters within the range allowed by observations, as summarized in \sec \ref{sec:input-param}. In \sec \ref{sec:cool} we briefly discuss the role of radiative cooling while we describe the implications for the CR spectrum in such winds in \sec \ref{sec:spectrum}.\\
\\
The topology of the solutions of the hydrodynamic equations  \ref{eq:Hcal-mass}-\ref{eq:Hcal-GAeff} depends on the nature of the critical points of the wind equation \ref{eq:Hcal-wind eq} (see \cite{Breitschwerdt-1991A&A...245...79B}; \citetalias{Recchia-2016-08082016}), i.e of the points in which the velocity derivative has zero numerator $(c_{*}^2= \frac{d\Phi}{dz}/\frac{1}{A} \frac{dA}{dz})$  and/or zero denominator ($u^2=c_{*}^2$). The point for which both the numerator and the denominator vanish is the critical (sonic) point, and it corresponds to the location where the flow velocity equals the compound sound speed, i.e $u=c_{*}$. The solution relevant for our problem is the one with velocity that starts subsonic at the wind base $z_0$, increases with $z$, goes through the critical point where it becomes supersonic and keeps increasing (wind acceleration). The wind launching velocity $u_0$ for this solution is found by imposing crossing through the critical point. For  launching velocities smaller than $u_0$, the flow remains subsonic everywhere and there is a point in which the numerator of \eq \ref{eq:Hcal-wind eq} vanishes. Such solutions are called ``breezes". For  launching velocities larger than $u_0$ but still subsonic, there is a point in which the denominator of \eq \ref{eq:Hcal-wind eq} vanishes and the corresponding solutions are unphysical.\\  
The hydrodynamic equations, and in particular the mass and energy conservation equations and the wind equation,
\begin{align}
& \rho u A=\textnormal{const} \label{eq:H2-mass} \\
&\frac{u^2}{2} +\frac{\gamma_g}{\gamma_g -1} \frac{P_g}{\rho} + \Phi +\frac{\gamma_{eff}}{\gamma_{eff} -1}  \frac{P_c}{\rho}\frac{u+v_A}{u} = \textnormal{const} \label{eq:H2-energy}\\
& \frac{du}{dz}=u\frac{c_{*}^2\, a(z) - g(z)}{u^2-c_{*}^2}, \label{eq:H2-wind eq}
\end{align}
(with reference to \eq \ref{eq:Hcal-wind eq} we define $a(z) \equiv \frac{1}{A}\frac{dA}{dz}$ and $g(z) \equiv \frac{d\Phi}{dz}$),
present many similarities with the Solar wind and the De Laval nozzle problem. In analogy with the Solar wind, we define a wind ("coronal") base where the boundary conditions for the problem are assigned, and we look for a solution in which the flow experiences a smooth transition from subsonic to supersonic regime, as in the De Laval nozzle problem. As in the Parker model for the Solar wind (see \cite{Parker-1965SSRv....4..666P}) the only possible transonic solution is the one which passes through the critical point, namely the point where both the numerator and the denominator of the wind equation are zero. The similarity with the De Laval nozzle  is even more evident if we compare the formal expression for the wind equation reported above and the De Laval nozzle:
\begin{align}
& \frac{1}{u}\frac{du}{dz} = \frac{a(z) -\frac{g(z)}{c^{*2}}}{M^2 -1} \;\;\; (\textnormal{Wind equation}) \label{eq:dudz-wind} \\ 
\\ \nonumber
& \frac{1}{u}\frac{du}{dz} = \frac{\mathfrak{a}(z)}{M^2 -1} \;\;\; (\textnormal{De Laval nozzle equation})\label{eq:dudz-DeLaval}
\end{align}
\begin{figure}
	\centering
	\label{fig:DeLaval}
	\includegraphics[width=\linewidth]{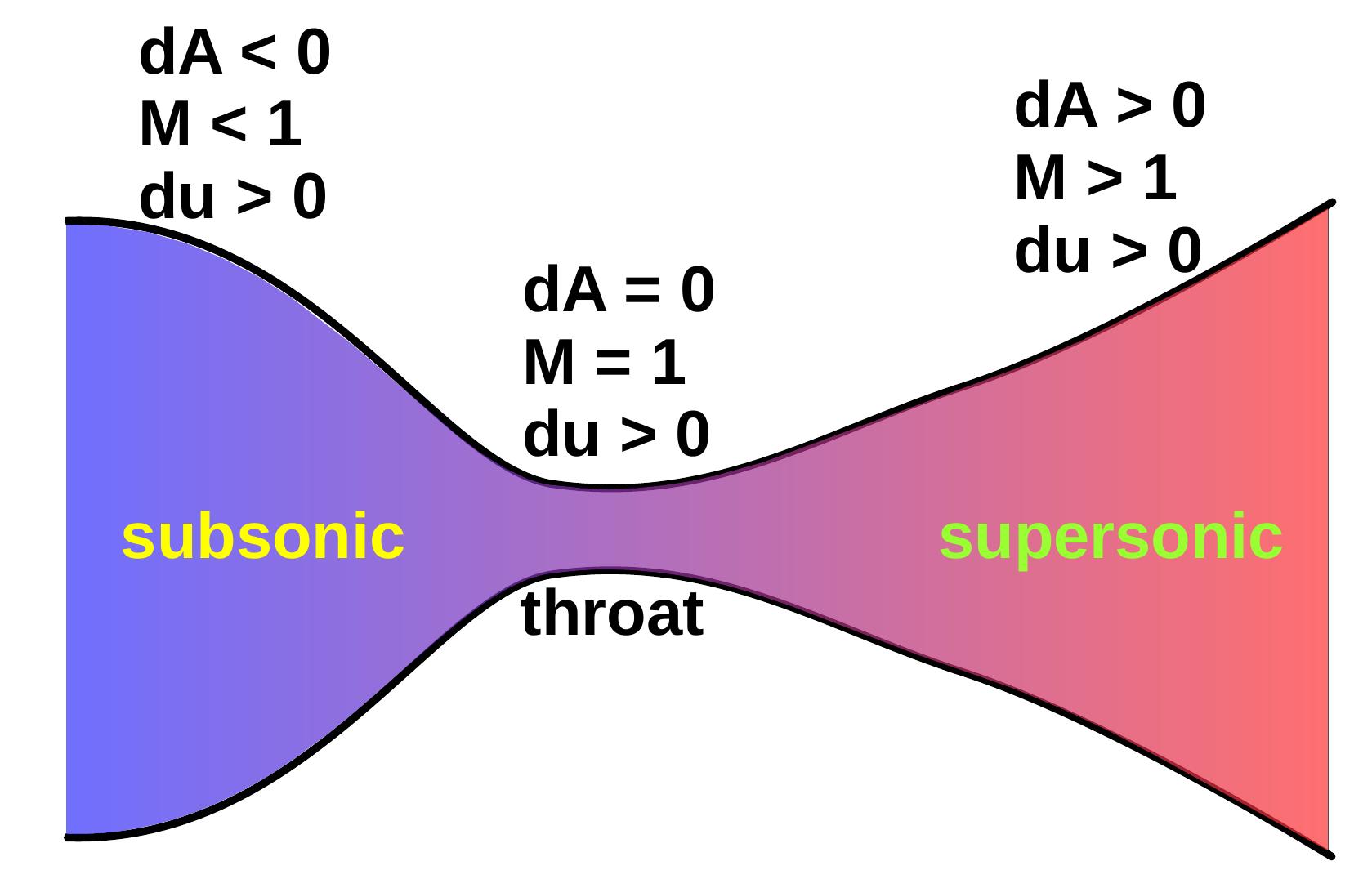}
	\caption{Scheme of a De Laval nozzle: the flow is smoothly accelerated (du/dz > 0 in all the duct) from the subsonic (M < 1) to the  supersonic regime (M > 1). The flow is subsonic in the converging section (dA/dz < 0), becomes sonic at the throat (dA/dz=0, M=1) and continues to accelerate to supersonic speed in the diverging section (dA/dz > 0).}
\end{figure}
where $M=u/c^{*}$ is the Mach number and $\mathfrak{a}(z) \equiv \frac{1}{A}\frac{dA}{dz}$ for the De Laval nozzle. In \fig \ref{fig:DeLaval} the scheme of a De Laval nozzle is shown. Notice that the nozzle has a converging and a diverging section. If we look at \eq \ref{eq:dudz-DeLaval}, we see that this converging-diverging duct makes a smooth subsonic-supersonic transition possible: the flow starts subsonic in the converging duct ($\mathfrak{a}(z) < 0$ and $M<1$), becomes sonic at the throat ($\mathfrak{a}(z) = 0$ and $M=1$) and is accelerated to supersonic speed in the diverging duct  ($\mathfrak{a}(z) > 0$ and $M >1$). In such configuration it is possible to constantly accelerate the flow along the nozzle ($du/dz$ remains positive). The sonic transition is only possible at the nozzle throat (choking), where $\mathfrak{a}=0$.\\
The numerator of the CR-driven wind equation \ref{eq:dudz-wind} presents an "effective area" term given by $a(z) -\frac{g(z)}{c^{*2}}$, which plays the same role of $\mathfrak{a}(z)$ in the De Laval equation. Thus, in order to have a smooth subsonic-supersonic transition, the effective area must have the same convergent-divergent behavior of the De Laval nozzle:
\begin{itemize}
	\item $u^2 < c^{*2}$ (subsonic regime)\\
	\\
	the gravitational term $g(z)$ must dominate in the numerator in order to have $du/dz > 0$. Thus, $g(z)$ is a "converging duct" term,\\
	
	\item $u^2 > c^{*2}$ (supersonic regime)\\
	\\
	the area term $a(z)$ must dominate in the numerator in order to have $du/dz > 0$. Thus, $a(z)$ is a "diverging duct" term.
\end{itemize}
It is clear from this analysis that if $a(z)=0$, i.e if the flux-tube area is constant, a wind solution cannot be achieved. Similarly, if the $a(z)>0$, namely the flow only expands (as in our wind model), the gravitational term $\frac{g(z)}{c^{*2}}$ allows for the presence of a stationary wind solution.\\
Finally, the passage through the critical point (choking condition for the De Laval nozzle), fixes the wind launching velocity $u_0$. Because all other magnitudes in \eq \ref{eq:H2-mass} and \ref{eq:H2-energy} are assigned at the wind base, this also fixes the mass and energy flux of the wind.\\  
\\
In order to carry out an analysis of the effects of the input parameters on the flow, it is helpful to define a reference model to which all other  cases can be compared. Because we are considering winds launched at the Sun position, we define our reference case starting from observations and we choose: $z_0 = 100$ pc for the wind base (winds launched near the Galactic disk), $n_0 = 6\times 10^{-3}$ cm$^{-3}$ for the gas density, $T_0 = 2\times 10^{6}$ K for the gas temperature, $P_{c0}= 4\times 10^{-13}$ erg/cm$^3$ for the CR pressure, $B_0 = 1\,\mu$G for the magnetic field, $Z_b = 15$ kpc and $\alpha = 2.0$ for the area parameters and NFW-Sofue  for the DM profile. The latter choice is motivated by the fact that the NFW profile is one of the most commonly used models for dark matter halos.\\
\begin{figure}
	\makebox[\linewidth][c]{%
		\subfigure[Reference case: velocities.\label{fig:ref-u}]%
		{\includegraphics[width=\columnwidth]{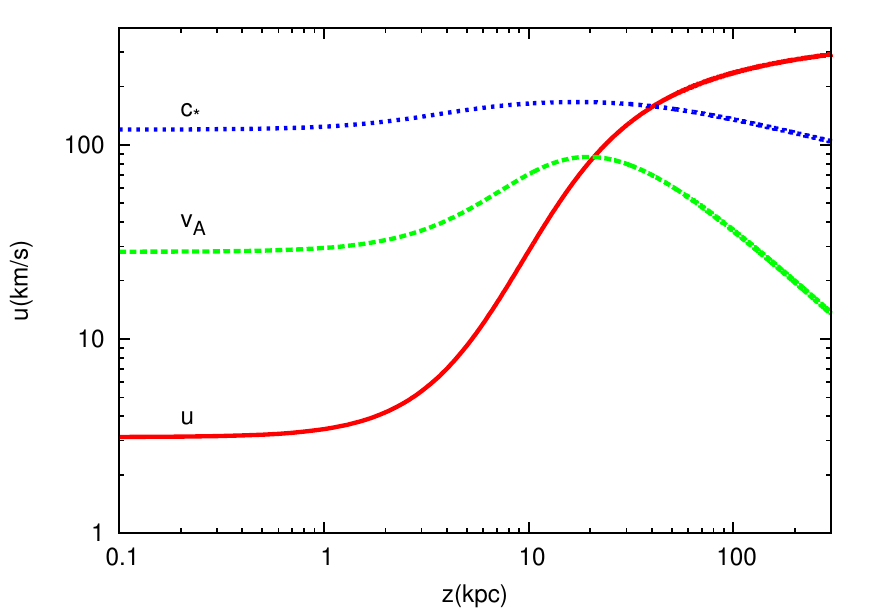}}
	}\\
	\makebox[\linewidth][c]{%
		\subfigure[Reference case: gas density and temperature.\label{fig:ref-n}]%
		{\includegraphics[width= \columnwidth]{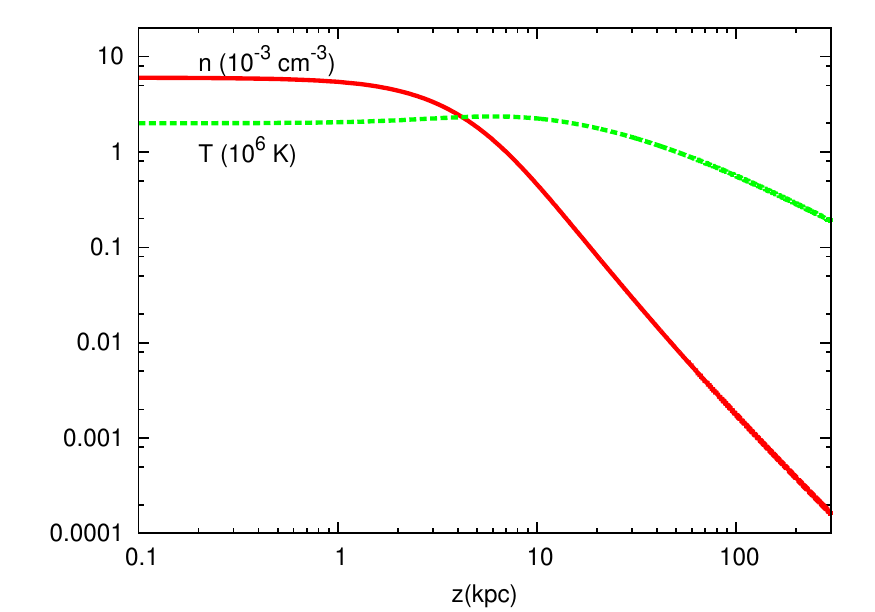}}
	}\\
	\makebox[\linewidth][c]{%
		\subfigure[Reference case: pressures.\label{fig:ref-P}]%
		{\includegraphics[width=\columnwidth]{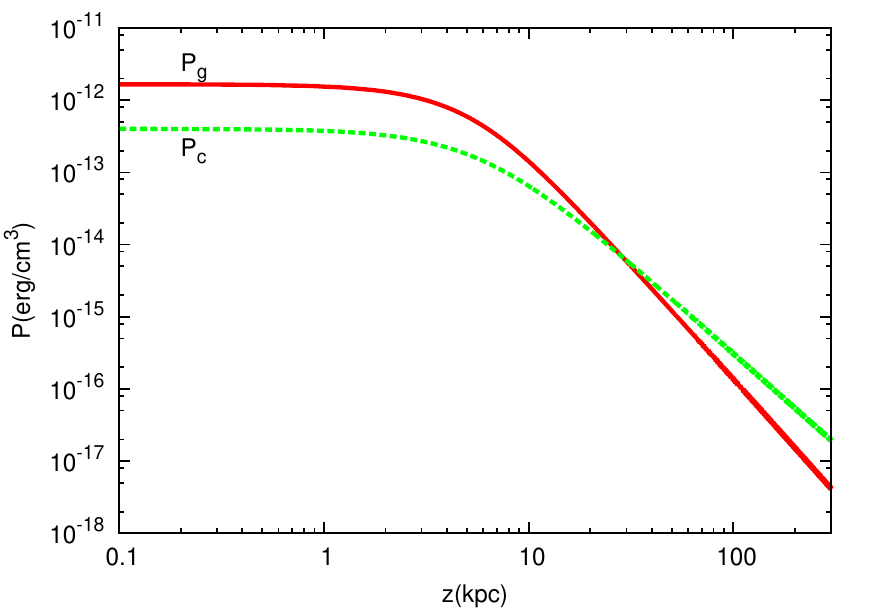}}
	}
	\caption{Reference case: a) wind, Alfv\'{e}n and compound sound speed; b) gas density and temperature; c) gas and CR pressures. The input parameters are: $z_0=100$ pc, $n_0 = 6\times 10^{-3}$ cm$^{-3}$, $T_0 = 2\times 10^{6}$ K, $P_{c0}= 4\times 10^{-13}$ erg/cm$^3$, $B_0 = 1\,\mu$G, $Z_b = 15$ kpc, $\alpha = 2.0$ and the NFW-Sofue dark matter profile.\label{fig:ref-case-param-study}}
\end{figure}
In \fig \ref{fig:ref-u}-\ref{fig:ref-P} we reported the spatial profile of the wind related quantities for the reference case, while the various components of the Galactic gravitational acceleration have been previously shown  in \fig \ref{fig:GravPot}. Notice that, unlike the bulge and disk components, which  die off at $\sim 10$ kpc, the DM contribution to the acceleration remains important out to the virial radius ($\sim$ 240 kpc for the NFW-Sofue).

For the reference case, at the wind base the flow is subsonic and sub-Alfv\'{e}nic and most of the work on the gas is due to the thermal pressure, which is four times larger than the CR pressure. CRs contribute to push the gas, however a considerable fraction of their energy is used to generate Alfv\'{e}n waves, which are quickly damped and heat the gas. This can be seen in \fig \ref{fig:ref-n}, where it is shown that the gas temperature increases up to $\sim 10$ kpc and then decreases in $z$ much slower than the gas density. Because of the CR-induced gas heating, also the gas pressure falls off in $z$ slower than it would in the absence of wave damping. Nevertheless the gas pressure decreases with height much faster than the CR pressure, and at $\sim 20$ kpc the CR pressure starts to dominate over the gas pressure. This is due to the fact that CRs have a smaller adiabatic index compared to the gas ($\gamma_c \sim 4/3$; $\gamma_g = 5/3$), which has important consequences on the relative contribution of the thermal and CR pressure to the wind launching. For instance, while the thermal pressure is more efficient in accelerating the gas near the wind base, the CR pressure is able to push the gas also at large $z$, often making an outflow possible where the thermal pressure alone would have failed (see \cite{Ipavich-1975p3566}; \cite{Breitschwerdt-1991A&A...245...79B}; \cite{Everett-2008p3580}).

Having defined a reference set of parameters, below we study the dependence of the outflow properties on the input parameters at the wind base. In \sec \ref{sec:n-T-Pc} we focus on the effect of changes of the gas density, gas temperature and CR pressure, while the analysis of the dependence on the area parameters is reported in \sec \ref{sec:area-param}.
\subsection{Effect of the gas density and temperature and of the Cosmic Ray pressure}
\label{sec:n-T-Pc}

In order to emphasize the dependence of the results on the input parameters, here we also include in our analysis parameters' values that are not strictly compatible with observations at the position of the Sun (discussed in \sec \ref{sec:input-param}).

\begin{table}
	\centering
	\caption {Dependence of the wind properties on the gas density. The gas temperature ($2\times 10^6$ K) and CR pressure ($4\times 10^{-13} \rm erg/cm^3$) are fixed. \.{m} is the mass loss rate of the wind.} \label{tab:n-depen} 
	\hspace{2.5pt}
	\begin{tabular}{|c||c|c|c|c|}
		\hline
		& &  & &\\ 
		Gas density 	& $u_0$  & $z_c$  & $u_f$  & \.{m}  \\
		$n_0^R = 6\times 10^{-3}$&  &  & & $4.7\times 10^{-4}$ \\ 
		$\rm (cm^{-3})$& (km/s) & (kpc)  & (km/s) & $(\rm M_{\odot} kpc^{-2} \rm yr^{-1})$\\ 
		& &  & &\\ 
		\hline
		0.2	& 39.7 & 18.9 & 595 & 2.5  \\ 
		\hline
		0.5	& 10.7 & 30.9 & 428 & 1.7 \\ 
		\hline
		1.0	& 3.1 & 40.6 & 403.7 & 1.0 \\ 
		\hline
		3.0	& 0.8 & 72.2 & 296 & 0.73 \\ 
		\hline
		6.0	& 0.3  & 96.5 & 269 & 0.53 \\
		\hline
	\end{tabular}
\end{table}
\begin{table}
	\caption {Dependence of the wind properties on the gas temperature. The gas density ($6\times 10^{-3} \rm cm^{-3}$) and CR pressure ($4\times 10^{-13} \rm erg/cm^3$) are fixed.} \label{tab:T-depen} 
	\begin{center}
		\begin{tabular}{|c||c|c|c|c|}
			\hline
			& &  & &\\ 
			Gas temperature & $u_0$  & $z_c$  & $u_f$ &  \.{m}  \\
			$T_0^R = 2\times 10^6$&  &  & & $4.7\times 10^{-4}$ \\ 
			(K) & (km/s) & (kpc)  & (km/s) & $(\rm M_{\odot} kpc^{-2} \rm yr^{-1})$\\ 
			& &  & & \\
			\hline
			0.5	& 3.09 & 34.3 & 419 & 0.98\\ 
			\hline
			1.0	& 3.1 & 40.6 & 403.7& 1  \\ 
			\hline
			2.0	& 6.9 & 104.9 & 243 & 2.2 \\ 
			\hline
			3.0	& 17.5 & 367.6 & 152.4 & 5.6 \\ 
			\hline
			4.0	& 182  & 13.32 & 263 & 58 \\
			\hline
			5.0	& 280 & 7.8 & 441.3 & 89\\ 
			\hline
			6.0	& 362  & 4.8 & 573.5 & 116 \\
			\hline
		\end{tabular}
	\end{center}
\end{table}
\begin{table}
	\caption {Dependence of the wind properties on the CR pressure. The gas density ($6\times 10^{-3} \rm cm^{-3}$) and  temperature ($2\times 10^{6}$ K) are fixed.} \label{tab:Pc-depen} 
	\begin{center}
		\begin{tabular}{|c||c|c|c|c|}
			\hline
			& &  & &\\ 
			CR pressure & $u_0$  & $z_c$  & $u_f$ &  \.{m}  \\
			$P_{c0}^R = 4\times 10^{-13}$ &  &  & & $4.7\times 10^{-4}$ \\ 
			$ (\rm erg/cm^3)$& (km/s) & (kpc)  & (km/s) & $(\rm M_{\odot} kpc^{-2} \rm yr^{-1})$\\
			& &  & & \\
			\hline
			0.2	& 0.72 & 64.1 & 339 & 0.2  \\ 
			\hline
			0.5	& 1.84 & 51.2 & 351 & 0.6 \\ 
			\hline
			1.0	& 3.1 & 40.6 & 403.7 & 1 \\ 
			\hline
			2.0	& 8.6 & 36.9 & 376.5 & 2.8 \\ 
			\hline
			3.0	& 14.5 & 32.6 & 394 & 4.6 \\ 
			\hline
			4.0	& 21.8  & 29.2 & 415 & 7.0 \\
			\hline
			5.0	& 30.7 & 26.2 & 439.6 & 9.8 \\ 
			\hline
			6.0	& 41.1  & 23.6 & 573.5 & 13 \\
			\hline
		\end{tabular}
	\end{center}
\end{table}
%
In \tab \ref{tab:n-depen}-\ref{tab:Pc-depen} we report the launching velocity $u_0$, the position of the sonic point $z_c$ and the terminal velocity $u_f$, for different values of the gas density (\tab \ref{tab:n-depen}), gas temperature (\tab \ref{tab:T-depen}) and CR pressure (\tab \ref{tab:Pc-depen}), as compared with the reference case of \fig \ref{fig:ref-case-param-study} (indicated by the superscript "R").

Some basic considerations can be put forward based on purely energetic grounds. Consider for instance what happens at fixed gas density and CR pressure values as in \tab \ref{tab:T-depen}: for temperatures below $\sim 0.4\, T_0^R$ it is impossible to launch a wind simply because the energy available is too small. In the opposite limit, for instance for temperature $\gtrsim 7\, T_0^R$, the gas is simply too hot and the outflow resembles more an evaporation process, which in reality would result in a non-stationary outflow that cannot be described using our formalism (see also \cite{Everett-2008p3580}).

At fixed gas temperature and CR pressure, an increase in the gas density leads to  a decrease of both the launching velocity $u_{0}$ and the terminal velocity $u_{f}$ (see \tab \ref{tab:n-depen}). This is not surprising, since a larger mass load makes more difficult for the pressure forces to launch a wind. Notice that the decrease of $u_0$ with increasing density is fast enough to drive down the mass loss rate of the wind (which is proportional to $n_0\, u_0$) despite the increasing density (see \tab \ref{tab:n-depen}).\\
As shown in \tab \ref{tab:T-depen} and \tab \ref{tab:Pc-depen}, increasing the temperature or the CR pressure at fixed gas density leads to an increase in the launching velocity (and because $n_0$ is fixed, also in the mass loss rate). Notice however that multiplying by factor of 6 the temperature and the CR pressure results in an increase of factor $\sim$ 100 and $\sim $14 in $u_0$ respectively. This fact can be explained keeping in mind that momentum and energy depositions before or after the sonic point affect the wind in different ways (see e.g \cite{lamers-1999}; \cite{Everett-2008p3580}): momentum and energy added before the critical point is reached can increase both the mass loss and the terminal velocity, while momentum and energy input after crossing the critical point can only affect the terminal velocity, since the mass loss is determined by the passage through the critical point. The gas adiabatic index (5/3) is larger than the CR adiabatic index ($ \sim $ 4/3), implying a larger gas pressure  gradient compared to the CR pressure gradient. Thus the thermal pressure is  more efficient than the CR pressure at increasing the mass loss.
In addition, notice that the  $z$ evolution of the CR pressure also depends on whether the flow is sub-Alfv\'enic or super-Alfv\'enic. In fact, with reference to \eq \ref{eq:Hcal-Pc}, we have the following limits for the CR pressure gradient (see \cite{Everett-2008p3580}):
\begin{align}
\lim_{u << v_A} \frac{dP_c}{dz} = \gamma_{eff}\frac{P_c}{2\rho} \frac{d\rho}{dz} \\
\nonumber \\
\lim_{u >> v_A} \frac{dP_c}{dz} = \gamma_{eff}\frac{P_c}{\rho} \frac{d\rho}{dz}.
\end{align}        
The effective CR adiabatic index is $\gamma_{eff}/2$ in the sub-Alfv\'enic regime and $\gamma_{eff}$ in the super-Alfv\'enic regime. As a consequence, CRs are less efficient at driving the wind in the sub-Alfv\'enic regime than in the super-Alfv\'enic. \\
The behavior of the terminal velocity with the gas temperature and the CR pressure is a bit more complicated being non monotonic. In \tab \ref{tab:T-depen} and \tab \ref{tab:Pc-depen} we can see that, when increasing both $P_{c0}$ and $\rho_0$, the terminal velocity in general does not increase. Keeping the gas density and the CR pressure at the reference values and increasing the temperature from $0.5\,T_0^R$ to $3\, T_0^R$, $u_f$ decreases by a  factor $\sim 3$, while from $3\,T_0^R$ to $6\, T_0^R$ it increases by factor of $\sim 4$. Similarly, keeping the gas density and temperature at the reference values and increasing the CR pressure from $0.2\,P_{c0}^R$ to $P_{c0}^R$, $u_f$ increases. It decreases when the CR pressure goes from  $P_{c0}^R$ to $2\, P_{c0}^R$, and finally increases for CR pressure from  $2\, P_{c0}^R$ to $6\, P_{c0}^R$.\\
In order to understand this behavior, notice that the terminal velocity is computed from the energy conservation equation \ref{eq:H2-energy}. At large $z$, where all other quantities drop to zero, all the energy density of the wind goes into kinetic energy of the gas, thus $u_f$ is related to the quantities at the wind base through
\begin{equation}\label{eq:H2-uf}
\frac{u_f^2}{2} =\frac{u_0^2}{2} +\frac{\gamma_g}{\gamma_g -1} \frac{P_{g0}}{\rho_0} + \Phi(z_0) +\frac{\gamma_{eff}}{\gamma_{eff} -1}  \frac{P_{c0}}{\rho_0}\frac{u_0+v_{A0}}{u_0} .
\end{equation}
Unlike the gas pressure term, which does not depend on $u_0$, the CR term depends on the Alfv\'{e}nic Mach number $M_{A0} = u_0/v_{A0}$ through the factor $(1+M_{A0})/M_{A0}$. This factor becomes large when $M_{A0}$ is small ($\ll 1$) and approaches unity when $M_{A0}$ gets large ($\gg 1$). Considering the results of \tab \ref{tab:T-depen}, we have that, starting from $0.5\, T_0^R$ and increasing the temperature, the launching velocity and the gas pressure increase, together with the kinetic and gas terms in \eq \ref{eq:H2-uf}. However, the CR term decreases, due to the factor $(1+M_{A0})/M_{A0}$. In the range $0.5\,T_0^R \,-\, 3\,T_0^R$ the decrease of the CR term dominates over the increase of the kinetic and gas terms, thus leading to a decrease of the terminal velocity with increasing temperature. At larger gas temperature the gas term finally starts to dominate over the CR term and the terminal velocity increases with the gas temperature. Notice that at large temperatures the wind is launched  super-Alfv\'{e}nic ($v_{A0} = 28$ km/s) and the CR term becomes practically independent of $M_{A0}$. The results of \tab \ref{tab:Pc-depen} can be explained with similar considerations.\\
The location of the sonic point is closer to the Galactic disk for smaller gas density and for larger gas temperature and CR pressures. This is due to the fact that in all three cases the gas is accelerated more easily and in general reaches the sonic point at smaller $z$.

Is is also worth stressing the role played by wave damping.  \cite{Breitschwerdt-1991A&A...245...79B} did not study the effect of wave damping in detail and the paper mainly focused on a  model where Alfv\'{e}n waves, produced through CR streaming instability, are not damped and can grow indefinitely. In such a model, the parameter space for the wind launching was much larger than in the case where wave damping is included, resulting in the possibility to easily launch winds also at relatively low temperatures and high gas densities. In the presence of wave damping the parameter space is reduced, mainly due to two factors. First, if waves are not damped, their pressure contribution, which grows  to reach more or less the same magnitude as the CR pressure, can significantly help in  pushing the gas. Second when the waves are damped, the heat input from wave damping increases the temperature (and thus the gas pressure) along $z$, which results in a smaller gas pressure gradient, i.e a smaller force acting on to the wind. If the contribution of wave damping is too intense, the gas heating can be so important so as to make the gas pressure increase with $z$, thus creating a stall in the outflow. In this situation the wind formation is prevented  (see \cite{Everett-2008p3580}).      

We conclude this overview with a discussion of the mass loss rate of the Galaxy due to winds. In Tables \ref{tab:n-depen}, \ref{tab:T-depen} and \ref{tab:Pc-depen} we report the mass loss rate per unit area for different values of the launching parameters of the wind. If we assume that such mass loss rate per unit area is the same as in the whole Galactic disk and we consider input parameters compatible with the halo observations,  we obtain  a  mass loss rate roughly in the range $\sim 0.5- 1.5 M_{\odot}/\rm yr$. Such values of the mass loss rate are of the same order of magnitude of the Galactic star formation rate ($\sim M_{\odot}$/yr; see e.g \cite{Robitaille-2010}), which means that CR-driven Galactic winds may play an important role in the evolution of the Milky Way.
\subsection{Effect of the flux-tube geometry}
\label{sec:area-param}
Here we analyze the impact of the flux-tube geometry through the variation of the two parameters $\alpha$ and $Z_b$ of \eq \ref{eq:area}, keeping the gas density, temperature and CR pressure to the reference values of \fig \ref{fig:ref-case-param-study} and using the NFW-Sofue DM profile.  The "mushroom" type geometry of \eq \ref{eq:area} reflects what we would intuitively expect for outflows from disk galaxies: the flow proceeds in nearly cylindrical symmetry up to about a distance $Z_b$ from the galactic disk and then opens up in a nearly spherical way.   Of course, in a full treatment of the CR-driven wind problem, the flow geometry should be calculated self-consistently by accounting for the large-scale magnetic field. However, such study is beyond the scope of the present work, hence we retain the geometry described by \eq \ref{eq:area} and we study how the wind properties get modified when $Z_b$ and $\alpha$ are changed.

The flux-tube geometry directly affects the acceleration of the outflow, as well as the gas density (see \eq \ref{eq:H2-mass} and \ref{eq:H2-wind eq}) and consequently the pressure gradients (see \eq \ref{eq:Hcal-Pg} and \ref{eq:Hcal-Pc}). Thus, it is not surprising that the shape chosen for the function $A(z)$ has important implications for the wind properties. In what follows we first discuss the effect produced by changing the length scale, $Z_b$, and then the effect of changing the exponent $\alpha$:

\subsubsection{Effect of changing the length scale $Z_b$}

\begin{figure}
	\makebox[\linewidth][c]{%
		\subfigure[Changing $Z_b$: velocities.\label{fig:Zb-u}]%
		{\includegraphics[width=\columnwidth]{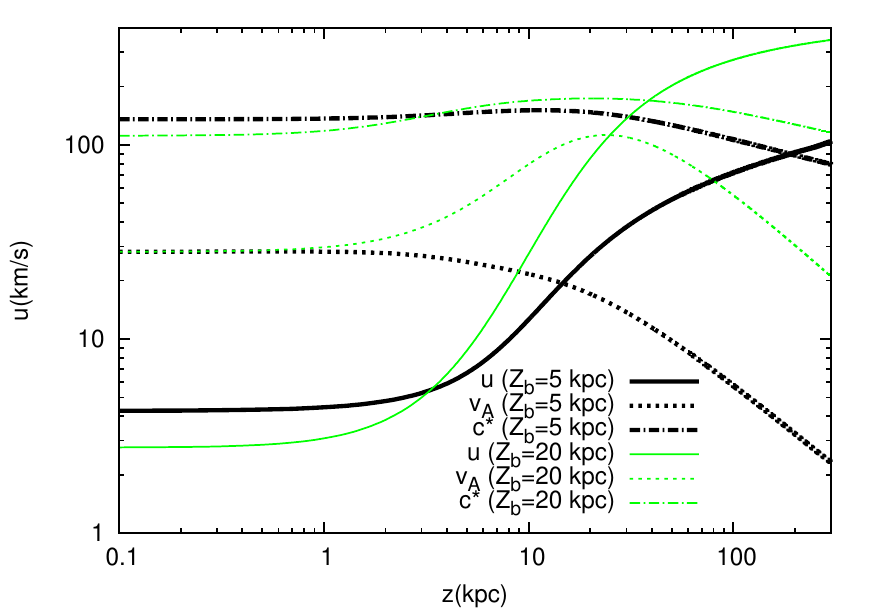}}
	}\\
	\makebox[\linewidth][c]{%
		\subfigure[Changing $Z_b$: gas density and temperature.\label{fig:Zb-n}]%
		{\includegraphics[width=\columnwidth]{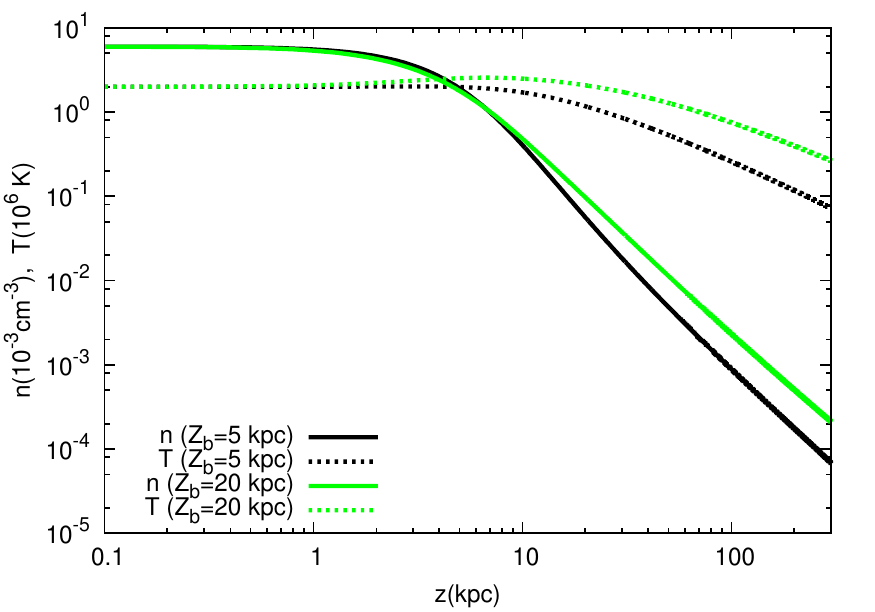}}
	}\\
	\makebox[\linewidth][c]{%
		\subfigure[Changing $Z_b$: pressures.\label{fig:Zb-P}]%
		{\includegraphics[width=\columnwidth]{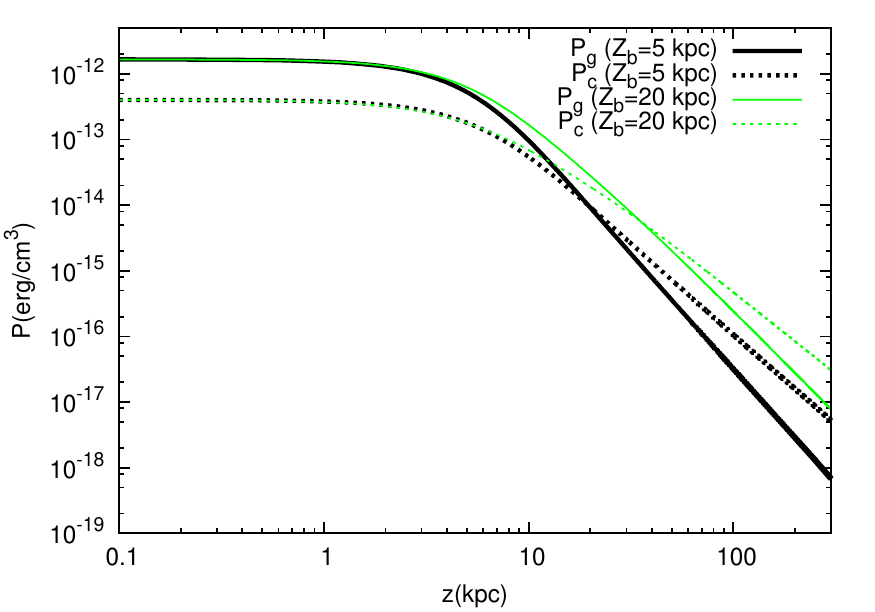}}
	}
	\caption{Comparison of wind profiles in case of changing the flux tube geometry. Two cases are shown,  $Z_b = 5$ kpc and $Z_b=20$ kpc, while the value of $\alpha$ is fixed to 2. The different plots are: a) wind, Alfv\'{e}n and compound sound speed; b) gas density and temperature; c) gas and CR pressure.\label{fig:Zb}}
\end{figure}

We fix $\alpha = 2.0$ and we change $Z_b$ in the range 5-20 kpc, stepped by 2.5 kpc. In \tab \ref{tab:Zb-depen} we show the wind launching velocity $u_0$, the wind terminal velocity $u_f$ and the location of the sonic point $z_c$ as functions of $Z_b$, while in \fig \ref{fig:Zb} we report the wind profile (relevant velocities, densities, temperature and pressures)  for the two extremal cases $Z_b = 5$ kpc and $Z_b=20$ kpc.
	
The plots in \fig \ref{fig:Zb} show that for smaller values of $Z_b$ vales the wind is launched with higher speed $u_0$, reaching the critical point $z_c$ farther away from the disc, but reaching a smaller final velocity $u_f$. These results can be explained as follows: for smaller values of $Z_b$, the adiabatic expansion of the gas associated to the geometric expansion of the outflow begins closer to the disc (smaller values of $z$), resulting in a larger density gradient, as shown in \fig \ref{fig:Zb-n}. Larger density gradients also correspond to larger pressure gradients (see \eq \ref{eq:Hcal-Pg} and \ref{eq:Hcal-Pc}) at smaller $z$. This results in an increase of the wind launching velocity (and, because the gas density is fixed, also in an enhanced mass loss rate).
	
Notice that, since $v_A = B/\sqrt{4\pi \rho}$ and $BA = const$ (see \eq \ref{eq:Hcal-B}), a steeper density profile also corresponds to a faster fall-off of the Alfv\'{e}n velocity with $z$ (see \fig \ref{fig:Zb-u}). For smaller values of $Z_b$ this leads to a flow which is typically "more super-Alfv\'{e}nic" along $z$, causing a more efficient driving of the CR wind (see discussion in \sec \ref{sec:n-T-Pc}), namely an extra contribution to the increase of $u_0$.
	
	\begin{table}
	\caption {Dependence of the wind properties on the area length scale $Z_b$. The gas density  ($3\times 10^{-3} \rm cm^{-3}$), gas temperature  ($2\times 10^6$ K) and CR pressure  ($4\times 10^{-13} \rm erg/cm^3$) are fixed. $\alpha = 2.0$.} \label{tab:Zb-depen} 
	\begin{center}
		\begin{tabular}{|c||c|c|c|}
			\hline
			& &  & \\ 
			$Z_b$ (kpc)	& $u_0$ (km/s) & $z_c$ (kpc)  & $u_f$ (km/s)  \\
			& &  & \\ 
			\hline
			5.0	& 4.3 & 189.9 & 218.4  \\ 
			\hline
			7.5	& 4.0 & 89.7 & 266.0 \\ 
			\hline
			10.0	& 3.7 & 56.4 & 315.7 \\ 
			\hline
			12.5	& 3.4 & 44.9 & 362.6 \\ 
			\hline
			15.0	& 3.1  & 40.6 & 403.7 \\
			\hline
			17.5	& 2.9 & 39.0 & 439.0 \\ 
			\hline
			20.0	& 2.8  & 38.6 & 469.1 \\
			\hline
		\end{tabular}
	\end{center}
	\end{table}
	Being all parameters at the wind base fixed, the increase of the wind launching velocity with decreasing $Z_b$ results in a decrease of the CR term and in an increase in the kinetic term in the energy conservation equation \ref{eq:H2-energy}. However, in the case considered, the first effect is dominant, and the terminal velocity decreases with decreasing $Z_b$. \\
	In the subsonic regime, the numerator of the wind equation \ref{eq:H2-wind eq} is dominated by the gravitational term. A smaller value of $Z_b$ makes the term $a(z)$ become important at smaller $z$ (the area starts to open  up at smaller $z$), while, as it can be seen in \fig \ref{fig:Zb-u},  $c^{*}$ depends weakly on $Z_b$. This results in a flatter velocity profile for smaller values of $Z_b$. The larger launching velocity at smaller $Z_b$ does not compensate the decrease in acceleration, leading to a sonic point located farther away from the disc. 

\subsubsection{Effect of changing the expansion index $\alpha$}

\begin{figure}
	\makebox[\linewidth][c]{%
		\subfigure[Changing $\alpha$: velocities.\label{fig:a-u}]%
		{\includegraphics[width=\columnwidth]{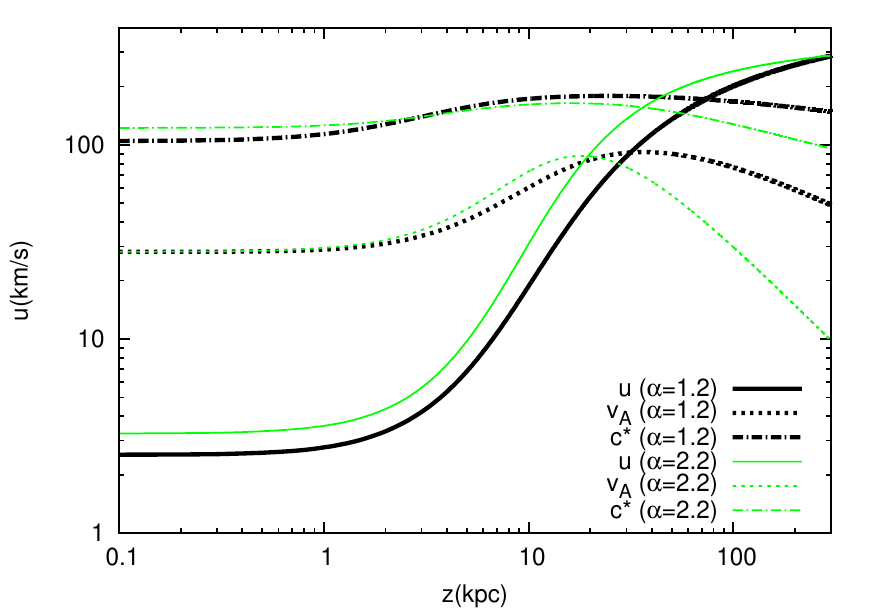}}
	}\\
	\makebox[\linewidth][c]{%
		\subfigure[Changing $\alpha$: gas density and temperature.\label{fig:a-n}]%
		{\includegraphics[width=\columnwidth]{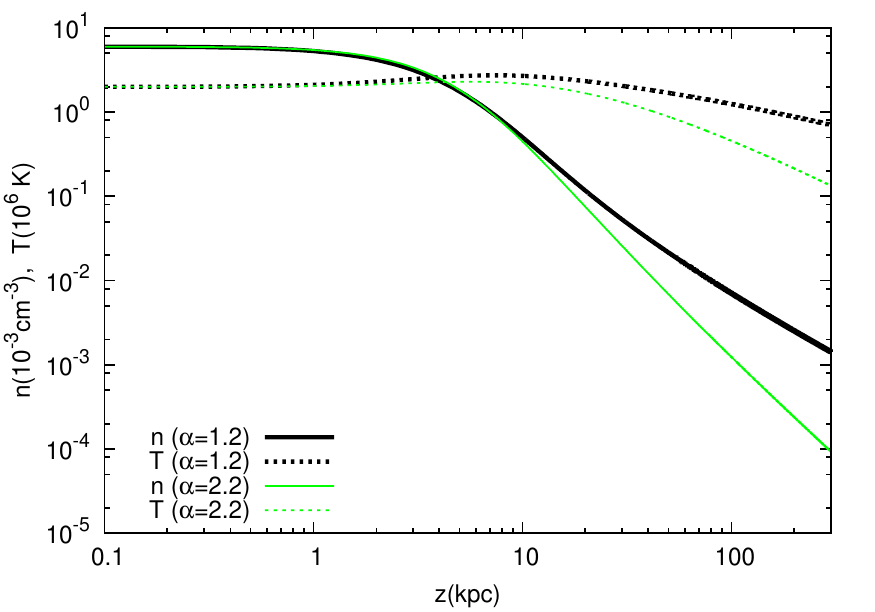}}
	}\\
	\makebox[\linewidth][c]{%
		\subfigure[Changing $\alpha$: pressures.\label{fig:a-P}]%
		{\includegraphics[width=\columnwidth]{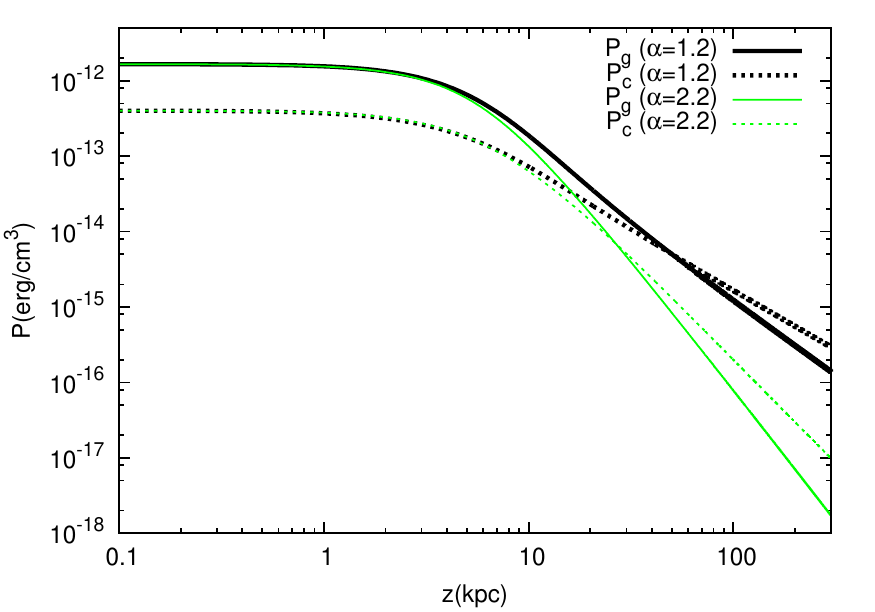}}
	}
	\caption{Comparison of wind profiles in case of changing the flux tube geometry. Two cases are shown,  $\alpha=1.2$ and $\alpha=2.2$, while the value of $Z_b$ is fixed to 15 kpc. The different plots are: a) wind, Alfv\'{e}n and compound sound speed; b) gas density and temperature; c) gas and CR pressures.\label{fig:a}}
\end{figure}

Here we fix $Z_b = 15$ kpc and we vary $\alpha$ in the range 1.2-2.2, stepped by 0.2.
In \tab \ref{tab:a-depen} we show the wind launching velocity $u_0$, the wind terminal velocity $u_f$ and the location of the sonic point $z_c$ as functions of $\alpha$. In  \fig \ref{fig:a} we show the wind properties only for the two extreme cases $\alpha = 1.2$ and $\alpha=2.2$.\\
	\begin{table}
	\caption {Dependence of the wind properties on the expansion index $\alpha$. The gas density  ($3\times 10^{-3} \rm cm^{-3}$), gas temperature  ($2\times 10^6$ K) and CR pressure  ($4\times 10^{-13} \rm erg/cm^3$) are fixed. $Z_b = 15$ kpc.} \label{tab:a-depen} 
	\begin{center}
		\begin{tabular}{|c||c|c|c|}
			\hline
			& &  & \\ 
			$\alpha$	& $u_0$ (km/s) & $z_c$ (kpc)  & $u_f$ (km/s)  \\
			& &  & \\ 
			\hline
			1.2	& 2.5 & 73.0 & 513.6  \\ 
			\hline
			1.4	& 2.7 & 61.2 & 482.0 \\ 
			\hline
			1.6 & 2.9 & 52.6 & 453.5 \\ 
			\hline
			1.8 & 3.0 & 45.9 & 427.6 \\ 
			\hline
			2.0 & 3.1  & 40.6 & 403.7 \\
			\hline
			2.2 & 3.3 & 36.2 & 382.0 \\ 
			\hline
		\end{tabular}
	\end{center}
	\end{table}
As it can be inferred from these plots, a larger value of $\alpha$ corresponds to smaller $z_c$, larger $u_0$ and smaller $u_f$. This behavior can be explained with considerations analogous to those invoked above for the case of a changing scale height $Z_b$. At $z > Z_b$, an increase in $\alpha$ corresponds to  steeper density profiles (see \fig \ref{fig:a-n}), i.e to larger pressure gradients, which act both in the subsonic and in the supersonic region (notice that in all cases $z_c > Z_b$). For this reason, when $\alpha$ increases, the launching velocity increases too (and consequently the mass loss rate increases), while the terminal velocity of the wind decreases.

Notice also that, as it can be seen in \fig \ref{fig:a-u}, the compound sound speed $c^*$ does not change appreciably when changing $\alpha$, especially in the subsonic regime. Moreover, for a given $Z_b$ the area of the flux tube $A(z)$ does not change much for $z<Z_b$. This leads to  velocity profiles which are practically parallel up to $\sim Z_b$ at all $\alpha$ (see \eq \ref{eq:H2-wind eq} and \fig \ref{fig:a-u}), and to a decrease of $z_c$ when  $\alpha$ increases. In fact, because the velocity profiles are nearly parallel within $\sim Z_b$ and $c^*$ is weakly dependent on $\alpha$ in the subsonic regime, at larger $u_0$ correspond smaller $z_c$. Finally, notice the steeper $c^*$ profile at $z > Z_b$  corresponding to larger values of $\alpha$. This is due to a more rapid fall off in $z$ of the density and pressures.
	
%
%
%
\subsection{Radiative cooling}
\label{sec:cool}

For gas temperature of order $\sim 10^6$ K radiative cooling may be important and affect in a potentially important manner the properties of CR driven winds or even the possibility to launch such winds. 

As shown by \cite{Dalgarno-1972}, in this range of temperature the ISM is cooled by the emission of forbidden lines and soft X-rays, at a  rate
\begin{equation}\label{eq:cool-T}
\frac{1}{T}\frac{d\,T}{dt} \sim \Lambda(T)n^2,
\end{equation}
where the cooling function is $\Lambda(\rm T\, \sim\, 10^6) \sim 2\times 10^{-23}$ erg cm$^3$/s for Solar metallicity. The dependence of the cooling rate as the square of the gas density suggests that the importance of the cooling will be much more important in the cases of dense winds. 

From the formal point of view, radiative cooling can be easily included in the wind hydrodynamic equations: \eq \ref{eq:Hcal-Pg} for the gas pressure becomes
\begin{equation}\label{eq:cool-Pg}
\frac{dP_g}{dz}=\gamma_g\frac{P_g}{\rho}\frac{d\rho}{dz}-(\gamma_g -1)\frac{v_A}{u}\frac{dP_c}{dz} - (\gamma_g -1)\Lambda(T)\frac{n^2}{u},
\end{equation}
while the wind equation \ref{eq:Hcal-wind eq} gets modified as
\begin{equation}\label{eq:cool-wind-eq}
\frac{1}{u}\frac{du}{dz}= \frac{c_{*}^2\, a(z) -g(z) + (\gamma_g -1)\Lambda(T)\frac{n}{u} }{u^2-c_{*}^2}.
\end{equation}
These expressions have been derived also by \cite{Breitschwerdt-1990ICRC....3..319D}, where it was also shown that the solution topology remains the same with the addition of radiative  cooling. Notice that the cooling term appears in the numerator of the wind equation \ref{eq:cool-wind-eq} with the same sign of the area term, namely it behaves as a "diverging duct" term in the De Laval nozzle picture discussed above.

Comparing the radiative cooling time and the wave heating time for the reference case of \fig \ref{fig:ref-case-param-study} ($n_0 = 6\times 10^{-3} \rm cm^{-3}$), we have $\tau_{cool} \sim 10^7$ yr and $\tau_{heat} \sim 10^8$ yr. This example shows that, in the absence of additional heating terms, the radiative cooling can be a quite fast process near the Galactic disk (where the gas density is larger) and can in principle prevent the wind launching. From the formal point of view, a large cooling term would result in a positive numerator at the wind base in  \eq \ref{eq:cool-wind-eq}, making a wind solution impossible.

On the other hand, the ISM is also heated by SN explosions through the injection of hot gas and magnetic turbulence that will eventually dissipate into thermal energy. Recently it was suggested that also Coulomb losses of CRs themselves might be a substantial source of heating (see \cite{Walker-2016}).
The relevance of cooling was already recognized by \cite{Breitschwerdt-1991A&A...245...79B} which, nevertheless, assumed that some sort of energy balance must exist between heating and cooling processes. It is currently unclear which mechanism may be responsible for sufficient heating to compensate cooling, but observations reveal that the temperature in the halo is indeed in the range $\sim 10^5-10^6$ K (see for instance \cite{Miller-2015-0004-637X-800-1-14}), hence supporting the idea that either cooling is negligible or it is balanced by one or more heating processes.

It is worth remarking that the cooling function in \eq \ref{eq:cool-T} has been estimated assuming that the halo has the same metallicity as the ISM in the solar vicinity, which might not be strictly true. In case of smaller metallicity, as one might expect for the gas in the halo, the cooling rate is somewhat reduced.

Given the lack of knowledge of detailed physical processes that may heat the base of the wind, here we adopted the same pragmatic attitude as \cite{Breitschwerdt-1991A&A...245...79B} and assume that radiative cooling is balanced by some heating process, so as to not include an explicit heating and cooling term in the wind equations (see also \citepalias{Recchia-2016-08082016}).

\section{Cosmic Ray spectrum}
\label{sec:spectrum}

As discussed in the sections above, the possibility to launch a CR driven wind and the properties of such winds depend upon the local conditions at the base of the wind, and the gravitational potential along the trajectory of the outflow. Of course, one of the important parameters of the problem is the CR pressure at the base of the wind. If we are interested in solving the problem at the position of the Sun we should recall that not only the CR pressure but also the CR spectrum is measured, so that the problem of launching a wind is much more constrained. In a non-linear theory, such as the one we adopt in this paper, the two quantities are not independent: the number of CRs at a given momentum determines the energy density of the self-produced waves and hence the diffusion coefficient, which in turn affects the spectrum and energy density of CRs. Moreover the same quantities have a direct effect on the wind properties, which also have a feedback on the CR spectrum through advection. Notice that in our iterative approach to solving the combined set of hydrodynamical equations for the wind and kinetic equation for CRs, the CR pressure is taken from observations ($P_{CR\,\odot} \sim 4 \times 10^{-13}$ erg/cm$^3$), but the spectrum is an output of the kinetic equation. Imposing that the predicted spectrum matches the observed flux at a given energy does not ensure that the spectrum also fits observations. In fact, it is in general the case that the spectrum corresponding to wind solutions is quite unlike the CR spectrum observed at the Earth (see also the discussion in \citetalias{Recchia-2016-08082016}).

This will be best discussed in \sec \ref{sec:ref-models-spectrum}, where we present some selected cases in which we solve the wind problem for input parameters suitable for the Sun's position in the Galaxy and we check the computed spectrum versus observations. We will show that typically the CR spectrum associated to wind solutions is harder than the observed one, as a result of a dominant role of advection of CRs with the wind. In some cases, the advection velocity can be so large as to dominate the CR transport up to energies of hundreds of GeV. A smaller Alfv\'{e}n speed (and possibly a smaller wind speed) is obtained by launching the wind with larger gas density (and/or at smaller magnetic field). On the other hand, it may happen that increasing the gas density at the base of the wind, so as to reduce the Alfv\'en speed and the wind speed, at some point the wind can no longer be launched. In some borderline cases, whether the wind is in fact launched or not may depend on the spatial distribution of dark matter, namely the contribution of dark matter to the gravitational potential. 

At CR energies $E\gtrsim 200$ GeV) the opening of the flux-tube area together with the energy dependence of the self-generated diffusion coefficient (see \eq \ref{eq:D self-gen}) makes the CR spectrum typically steeper than the observed one. However, as discussed in \citetalias{Recchia-2016-08082016}, the high energy behaviour of the CR flux may be considerably affected by the presence of non self-generated turbulence in the near disc region (within $\sim 1$ kpc from the base of the wind). In fact such turbulence should be postulated in this class of models if to account for the spectral hardening measured by the PAMELA \cite[]{pamhard} and AMS-02 \cite[]{amshard} experiments. 

In \sec \ref{sec:spectrum-obs-DM}, we discuss the impact of the DM halo profile on the CR spectrum,  
and we show that it is possible to find a wind model which is consistent with both the recent observations illustrated by \cite{Miller-2015-0004-637X-800-1-14}  and with the observed CR spectrum. 

\subsection{Reference models}
\label{sec:ref-models-spectrum}

\begin{figure}
	\makebox[\linewidth][c]{%
		\subfigure[Wind velocity, Alfv\'{e}n velocity and sound speed.\label{fig:modelA-u}]%
		{\includegraphics[width=\linewidth]{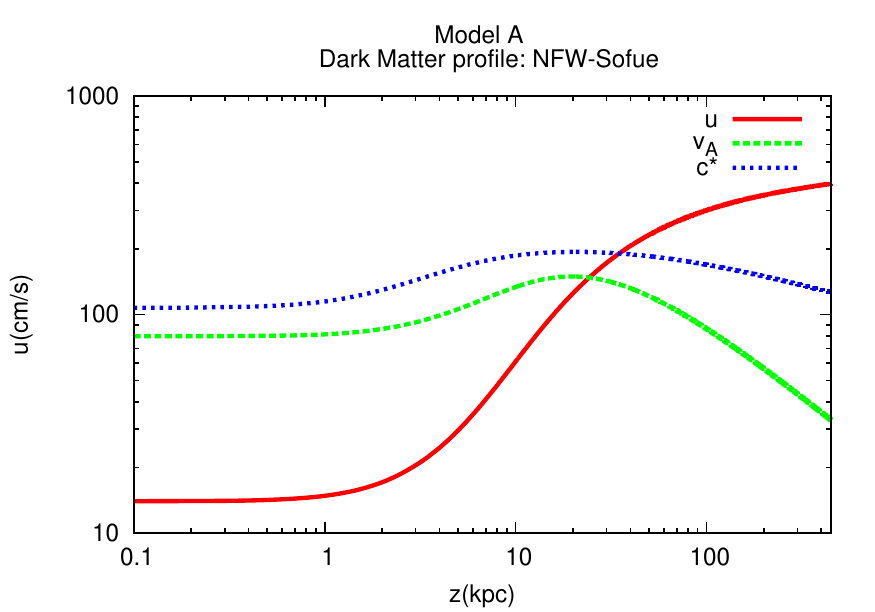}}
	}\\
	\makebox[\linewidth][c]{%
		
		\subfigure[Density in units of $10^{-3} \rm cm^{-3}$ and temperature in units of $10^6$ K.
		\label{fig:modelA-n}]%
		{\includegraphics[width=\linewidth]{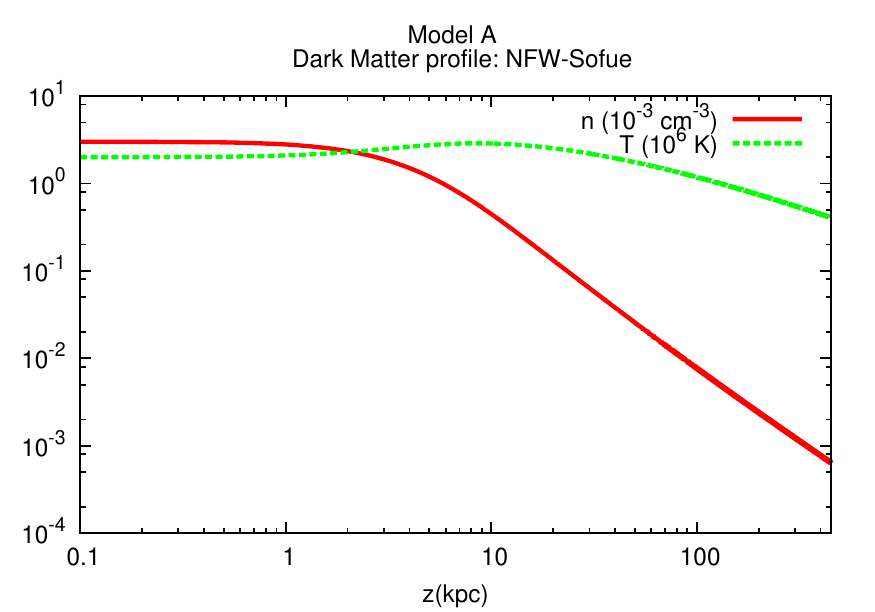}}
	}\\
	\makebox[\linewidth][c]{%
		\subfigure[Gas pressure, CR pressure obtained from the hydrodynamic and kinetic calculations,  wave pressure obtained from the CR transport equation. \label{fig:modelA-P}]%
		{\includegraphics[width=\linewidth]{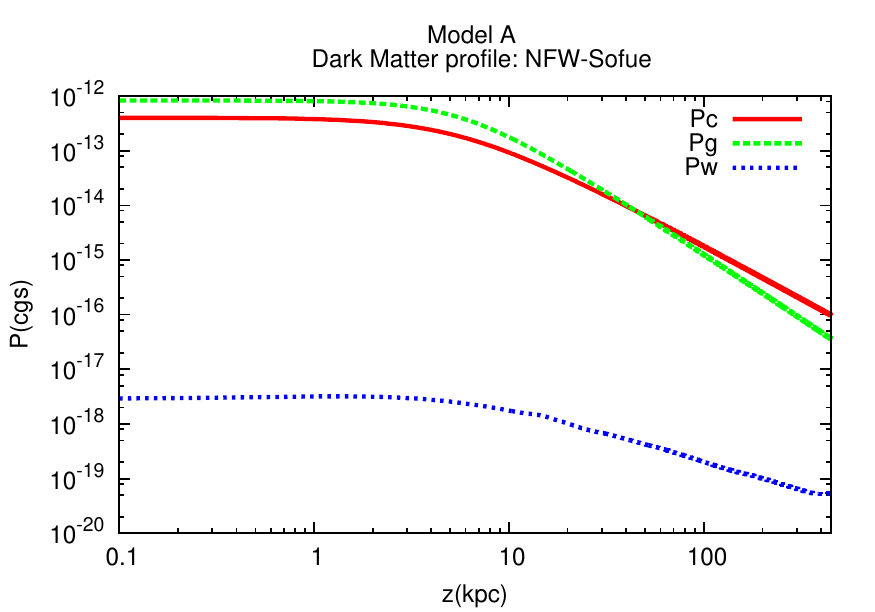}}
		
	}
	\caption{Results for Model-A: at the wind base, $z_0$, the gas density is $n_0 = 3\times 10^{-3} \rm cm^{-3}$ and the magnetic field is $B_0 = 2\mu$G. The CR injection spectrum has been normalized in order to get the observed CR pressure  $P_{c0}=4\times 10^{-13}~\rm erg/cm^{3}$ at $z_0$. \label{fig:modelA}}
\end{figure}

Here we discuss three scenarios in which a wind is launched at the position of the Sun (Galactocentric distance $R_{\odot} = 8.5$ kpc, $z_0= 100$ pc) with parameters that are compatible with observations. Following the discussion reported in \sec \ref{sec:input-param}, we fix the gas temperature as $T_0 = 2\times 10^{6}$ K and the parameters of the flux-tube  geometry as $Z_b=15$ kpc and $\alpha =1.5$. The calculations are carried out assuming that dark matter is distributed according to a NFW-Sofue profile and we fix the slope of the injection spectrum as $\gamma=4.3$. The three scenarios differ in the values of the gas density $n_0$ and magnetic field $B_0$ at the wind base and in the normalization of the CR injection spectrum (which is normalized either to reproduce the observed CR pressure or to match the computed and observed spectra at 50 GeV):
in the first scenario (Model A) we assume $n_{0}=0.003$ cm$^{-3}$ and $B_{0}=2\,\mu $G at the wind launching point $z_0$. In addition, we normalize the injection spectrum in order to get the observed CR pressure  $P_{c0}=4\times 10^{-13}~\rm erg/cm^{3}$ at $z_0$. This implies the following condition to be fulfilled: 
	\begin{equation*}
	\frac{\xi_{CR}}{0.1} \frac{\mathcal{R}_{SN}}{1/30 \rm \, yr^{-1}} \approx 1.1
	\end{equation*} 
for the injection term (\eq \ref{eq:injection}). The results for Model A are summarized in \fig \ref{fig:modelA}: \fig \ref{fig:modelA-u} shows the wind velocity, the Alfv\'{e}n speed and the compound sound speed, while wind density and temperature are shown in \fig \ref{fig:modelA-n}. The wind is launched sub-Alfv\'{e}nic with $u_0 = $ 14 km/s and becomes sonic at $z_c = 35$ kpc, while the Alfv\'{e}n speed at $z_0$ is 79 km/s. The CR pressure is shown in \fig \ref{fig:modelA-P} together with the gas pressure and the wave pressure. The latter  is computed from the equilibrium wave spectrum, obtained by equating the growth rate of waves due to CR streaming instability and the damping rate due to NLLD (see \eq \ref{eq:wave spectrum}). Notice that the wave pressure is much smaller than the gas and CR pressure, thereby justifying a posteriori the fact that it is neglected in the hydrodynamic equations. 

\begin{figure}
	\makebox[\linewidth][c]{%
		\subfigure[Wind velocity, Alfv\'{e}n velocity and sound speed.\label{fig:modelB-u}]%
		{\includegraphics[width=\linewidth]{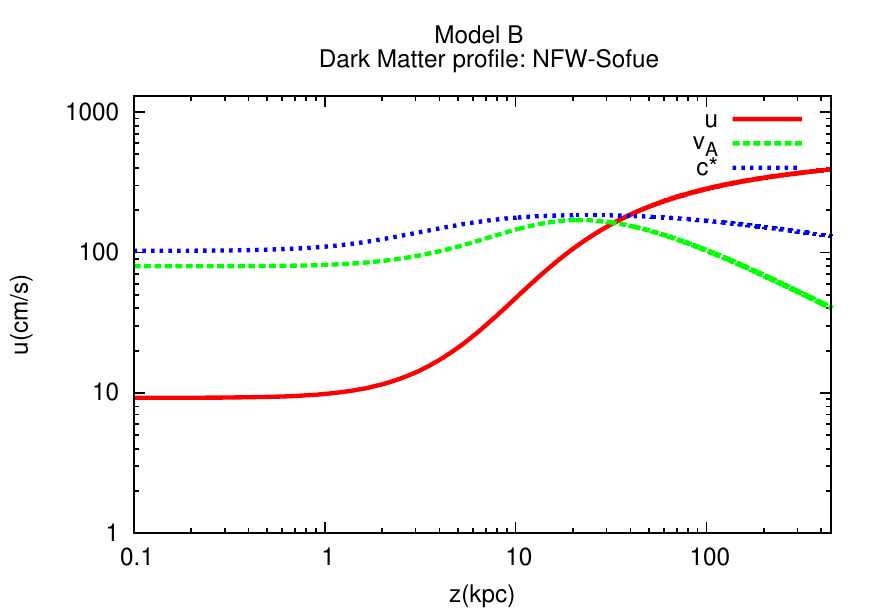}}

	}\\
	\makebox[\linewidth][c]{%
		
		\subfigure[Density in units of $10^{-3} \rm cm^{-3}$ and temperature in units of $10^6$ K.
		\label{fig:modelB-n}]%
		{\includegraphics[width=\linewidth]{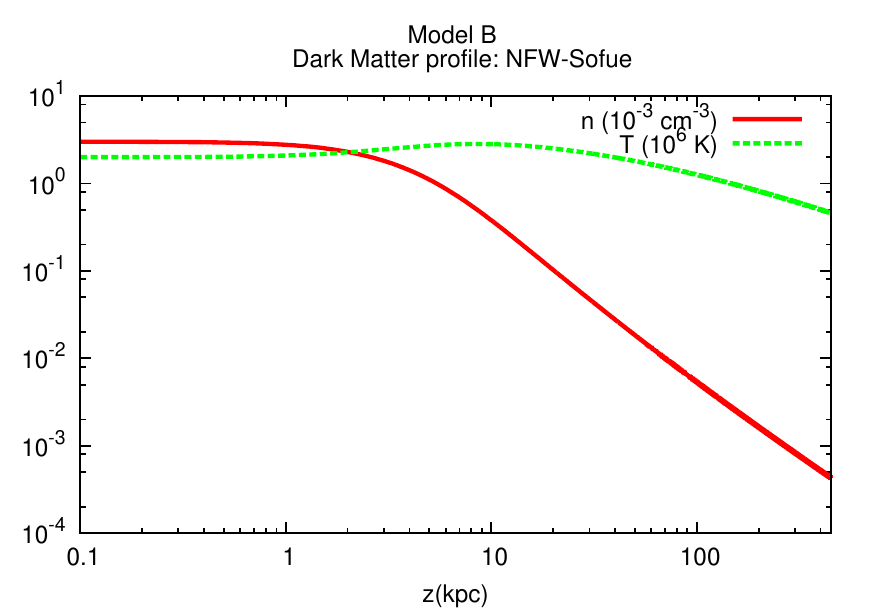}}
			
	}\\
	\makebox[\linewidth][c]{%
		\subfigure[Gas pressure, CR pressure obtained from the hydrodynamic and kinetic calculations,  wave pressure obtained from the CR transport equation. \label{fig:modelB-P}]%
		{\includegraphics[width=\linewidth]{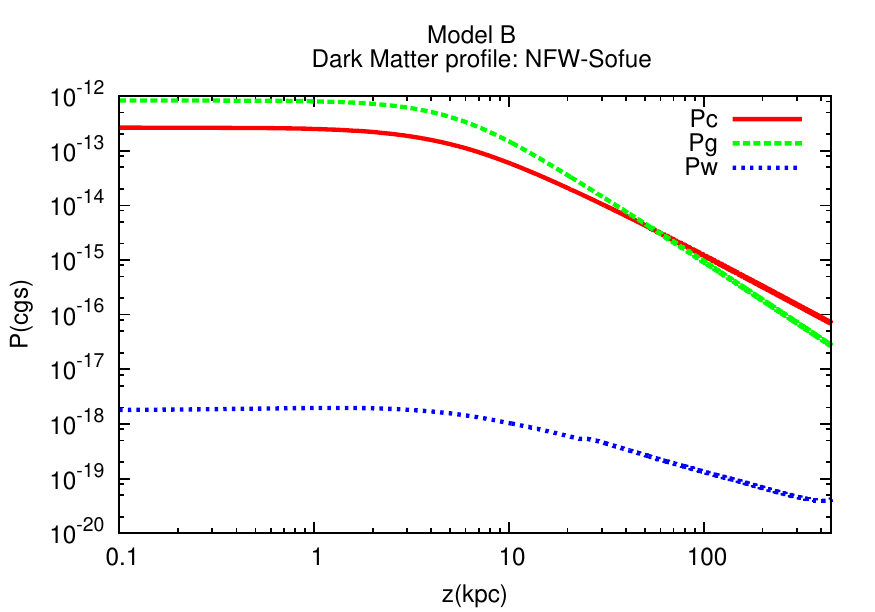}}
	}
	\caption{Results for Model-B: at the wind base, $z_0$, the gas density is $n_0 = 3\times 10^{-3} \rm cm^{-3}$ and the magnetic field is $B_0 = 2\mu$G. The CR injection spectrum has been normalized in order to reproduce the observed CR spectrum at 50 GeV. 
	\label{fig:modelB}}
\end{figure}

In the second model (Model B) we assume again a gas density $n_{0}=0.003$ cm$^{-3}$ and a magnetic field $B_{0}=2\,\mu $G at the base of the wind. However, we normalize the injection spectrum in a way that the computed flux of CR protons at the location of the Sun is the same as measured at an energy of $50$ GeV. This constraint implies the following condition on the injection parameters: 
	\begin{equation*}
	\frac{\xi_{CR}}{0.1} \frac{\mathcal{R}_{SN}}{1/30 \rm \, yr^{-1}} \approx 0.75.
	\end{equation*} 
The corresponding wind velocity, the Alfv\'{e}n speed and the compound sound speed are shown in \fig \ref{fig:modelB-u} while the wind density and temperature are shown in \fig \ref{fig:modelB-n}. The wind is launched sub-Alfv\'{e}nic with $u_0 = $ 9 km/s and becomes sonic at $z_c = 39$ kpc, while the Alfv\'{e}n speed at $z_0$ is 79 km/s. The CR pressure is shown in \fig \ref{fig:modelB-P}, together with the gas pressure and wave pressure. Notice that, because we are normalizing the spectrum to the observed proton flux at $50$ GeV rather than on the CR pressure at $z_0$, the latter is an output of the calculation, with value $P_{c0}=2.6\times 10^{-13}~\rm erg/cm^{3}$. This explains the difference in wind velocity between Model A and Model B.

In the third scenario (Model C), we assume a gas density $n_{0}=0.006$ cm$^{-3}$ and a magnetic field $B_{0}=1\,\mu $G at the base of the wind. As for Model B, we normalize the injection spectrum in order to reproduce the observed CR flux at 50 GeV, which implies: 
	\begin{equation*}
	\frac{\xi_{CR}}{0.1} \frac{\mathcal{R}_{SN}}{1/30 \rm \, yr^{-1}} \approx 0.32.
	\end{equation*} 

\begin{figure}
	\makebox[\linewidth][c]{%
		\subfigure[Wind velocity, Alfv\'{e}n velocity and sound speed.\label{fig:modelC-u}]%
		{\includegraphics[width=\linewidth]{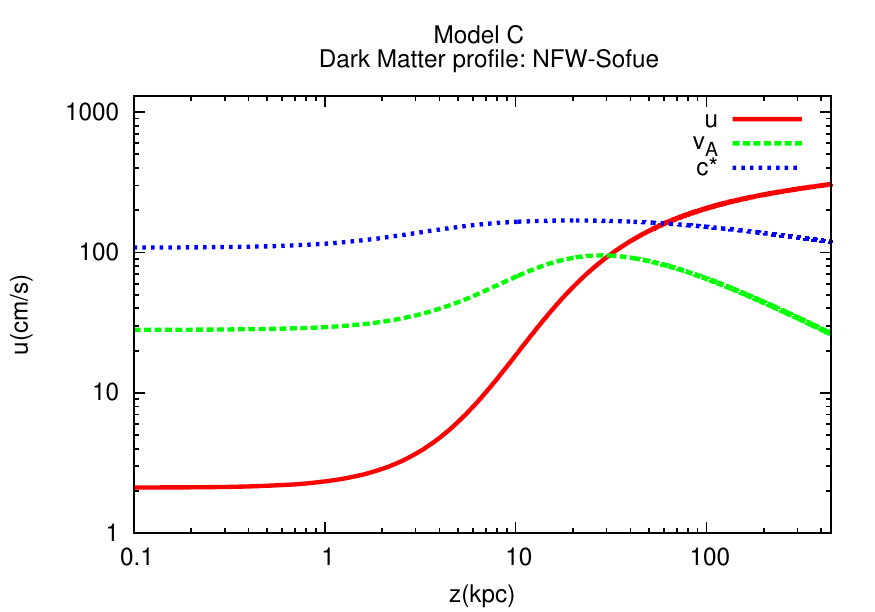}}
	}\\
	\makebox[\linewidth][c]{%
		\subfigure[Density in units of $10^{-3} \rm cm^{-3}$ and temperature in units of $10^6$ K.
		\label{fig:modelC-n}]%
		{\includegraphics[width=\linewidth]{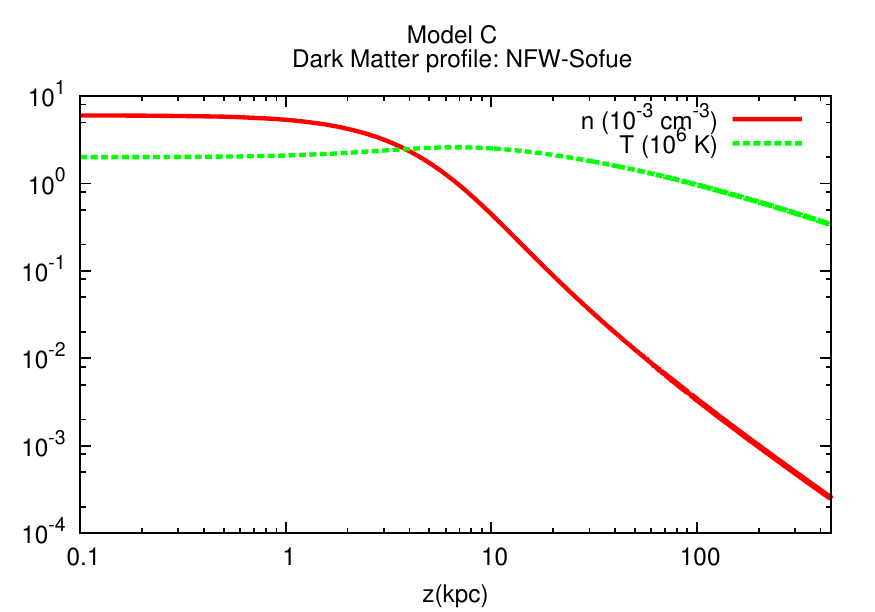}}
	}\\
	\makebox[\linewidth][c]{%
		\subfigure[Gas pressure, CR pressure obtained from the hydrodynamic and kinetic calculations,  wave pressure obtained from the CR transport equation. \label{fig:modelC-P}]%
		{\includegraphics[width=\linewidth]{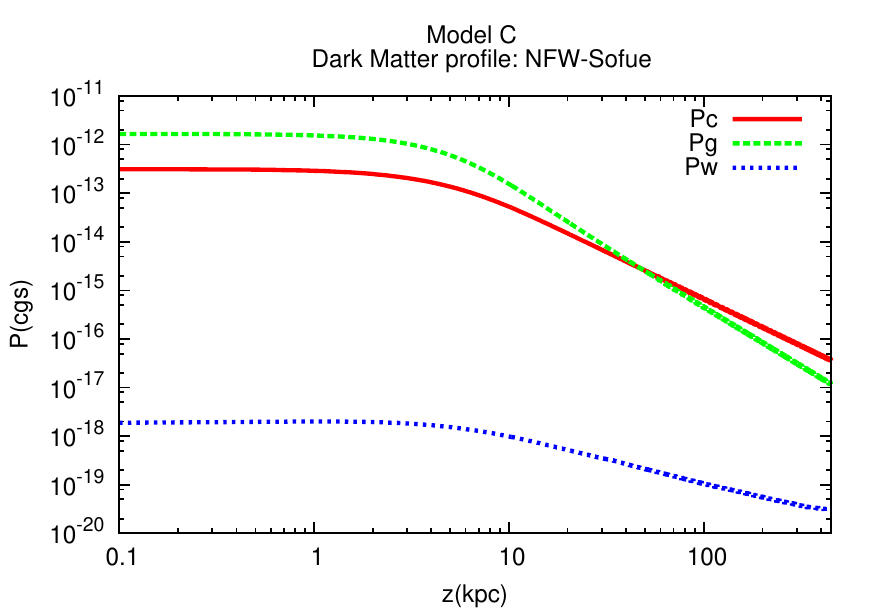}}
	}
	\caption{Results for Model-C: at the wind base, $z_0$, the gas density is $n_0 = 6\times 10^{-3} \rm cm^{-3}$ and the magnetic field is $B_0 = 1\mu$G. The CR injection spectrum has been normalized in order to reproduce the observed CR spectrum at 50 GeV. 
	\label{fig:modelC}}
\end{figure}

The wind velocity, the Alfv\'{e}n speed and the compound sound speed for Model C are shown in \fig \ref{fig:modelC-u}, while \fig \ref{fig:modelC-n} illustrates the evolution of the wind density and temperature. The wind is launched sub-Alfv\'{e}nic with $u_0 = 2$ km/s and becomes sonic at $z_c = 60$ kpc, while the Alfv\'{e}n speed at $z_0$ is 28 km/s. The CR pressure is shown in \fig \ref{fig:modelC-P}, together with the gas and the wave pressure. As for Model B, the CR pressure at $z_0$ is an output of the calculation, $P_{c0}=3.2\times 10^{-13}~\rm erg/cm^{3}$.

%
%
%
%

\begin{figure}
	\makebox[\linewidth][c]{%
		\subfigure[Model-A: CR spectrum.
		\label{fig:modelA-spectrum}]%
		{\includegraphics[width=\linewidth]{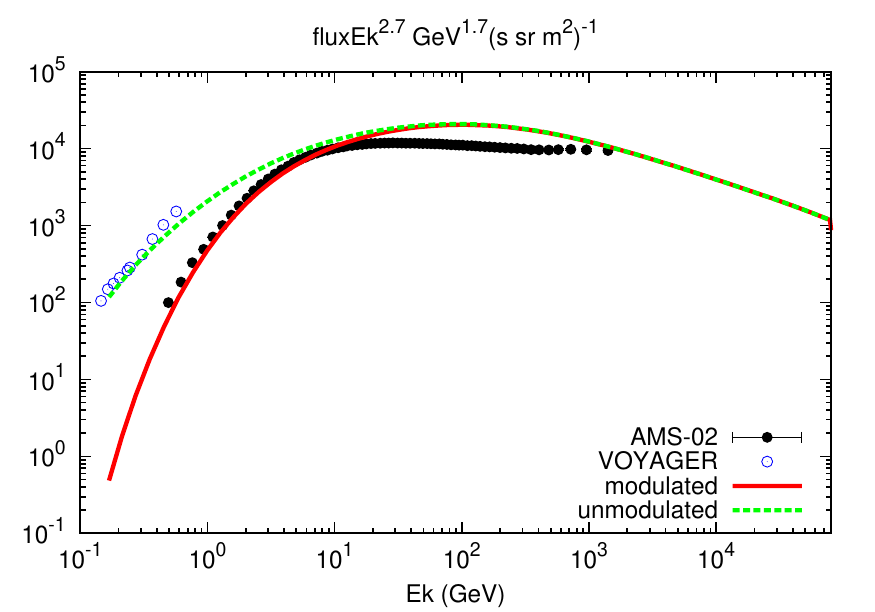}}
	}\\
	\makebox[\linewidth][c]{%
		\subfigure[Model-B: CR spectrum.
		\label{fig:modelB-spectrum}]
		{\includegraphics[width=\linewidth]{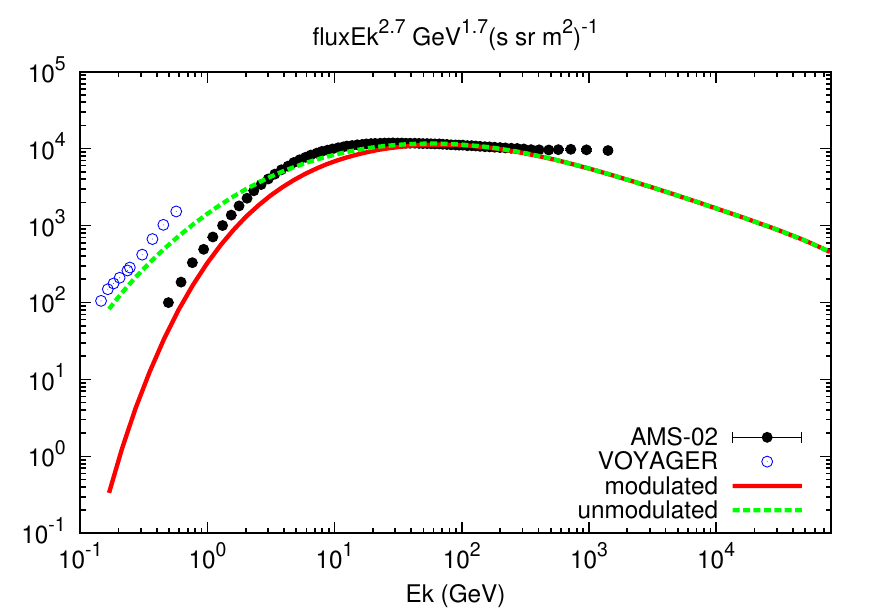}}
	}\\
	\makebox[\linewidth][c]{%
		\subfigure[Model-C: CR spectrum.
		\label{fig:modelC-spectrum}]
		{\includegraphics[width=\linewidth]{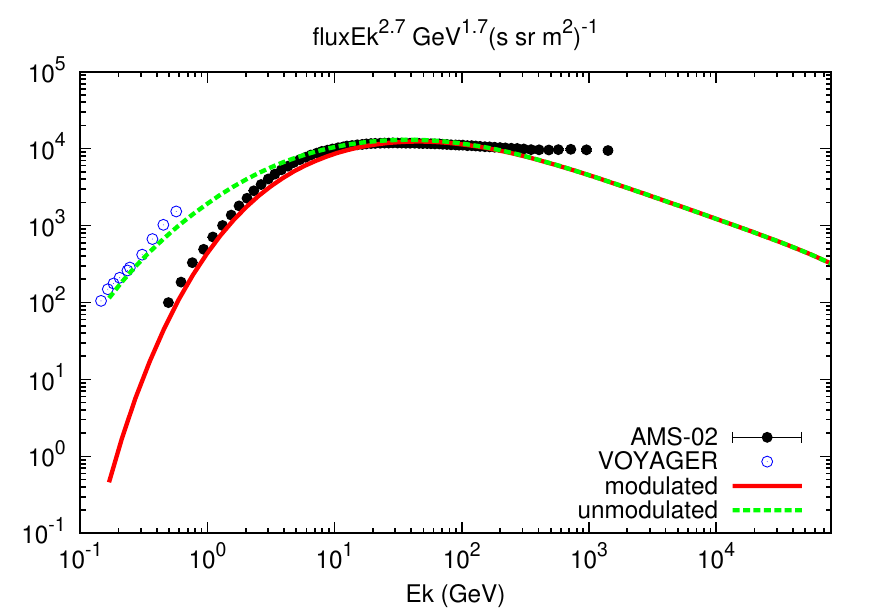}}
	}
	\caption{CR spectrum compared to the VOYAGER and AMS-02 data for Model A, B and C. Solar modulation has been applied with $\Phi = $500 MV.
	\label{fig:spectraA-B-C}}
\end{figure}

The spectrum of CR protons at the position of the Sun in the three scenarios discussed above is shown in Figs. \ref{fig:modelA-spectrum}, \ref{fig:modelB-spectrum} anf \ref{fig:modelC-spectrum} compared with the proton flux as measured by the VOYAGER-I \cite[]{Voyager}  at low energies and by AMS-02 \cite[]{amshard} at higher energies. The red curve illustrates the effect of solar modulation, with the simple recipe of \cite{DiBernardo-2010}. One should  not overestimate the importance of the comparison of our predicted CR spectrum with observations at energies below $\sim 1$ GeV, because in such energy region energy losses become important and such effects have not been included in the calculations presented in this paper. 

In all the three cases introduced above, the CR advection velocity near the wind base is dominated by the Alfv\'{e}n speed. The advection velocity at the wind base is $u_0 + v_{A0} \sim 93$ km/s, $\sim 88$ km/s and $\sim 30$ km/s for model-A, B and C respectively. An important feature of the spectra in the disk, shown for the three models in \fig \ref{fig:spectraA-B-C}, is the presence of hard spectra which extend to higher energies for increasing advection velocity. This is due to the fact that, the larger the CR advection velocity, the higher the CR energy at which advection dominates upon diffusion in the CR transport equation, thus producing a spectrum with spectral index close to the injection spectral index.

From \fig \ref{fig:modelC-spectrum}, one can see that below $\sim 200$ GeV Model-C resemble the observed CR spectrum, while the harder spectra in Model A and B appear to be quite unlike the observed one, even qualitatively. This is a rather general conclusion: even using values for the environmental parameters that are compatible with observations (see \sec \ref{sec:param-study}) and that lead to wind launching, the corresponding CR spectrum at the Sun's position is, in general, qualitatively different from the observed one, even if the total CR pressure may be close to the measured one. 

Making use of the points discussed in \sec \ref{sec:param-study} concerning the dependence of the wind properties on the launching parameters, we can infer what would happen by changing the gas density and temperature and the magnetic field at the wind base (always within the observational constraints). By increasing the magnetic field and/or decreasing the gas density the Alfv\'{e}n speed would increase and, consequently, harder spectra would be inferred. In addition,  a decrease of the gas density, as well as an increase of the gas temperature, leads to an increase of the wind launching velocity, with similar implications for the CR spectrum. On the other hand, an increase in the gas density would lead to smaller advection velocities (decreasing both the Alfv\'{e}n and the wind speed), hence a generally steeper spectrum. However, when the density becomes too high it may become impossible to launch a wind. The same considerations hold when decreasing the gas temperature: the Alfv\'{e}n speed is not affected but the wind velocity decreases. This situation is particularly important for winds in which the launching velocity is comparable or larger than the Alfv\'{e}n speed. Also in this case, for smaller temperatures the wind launching may be inhibited.
%
%
%
%

\subsection{Models versus observations: the effect of the Dark Matter halo}
\label{sec:spectrum-obs-DM}

Here we present a wind model whose parameters are compatible with properties of the Galactic halo derived in \cite{Miller-2015-0004-637X-800-1-14}, i.e Model-C of \sec \ref{sec:ref-models-spectrum} (see discussion below), and we study how the CR spectrum depends on the DM halo profile. In particular, we show that it is possible to have a good agreement with the observed CR proton spectrum, depending on the choice of the DM halo potential.
\begin{figure}
	\makebox[\linewidth][c]{%
		\subfigure[Wind and compound sound speed. \label{fig:s-DM-u-cs}]%
		{\includegraphics[width=\linewidth]{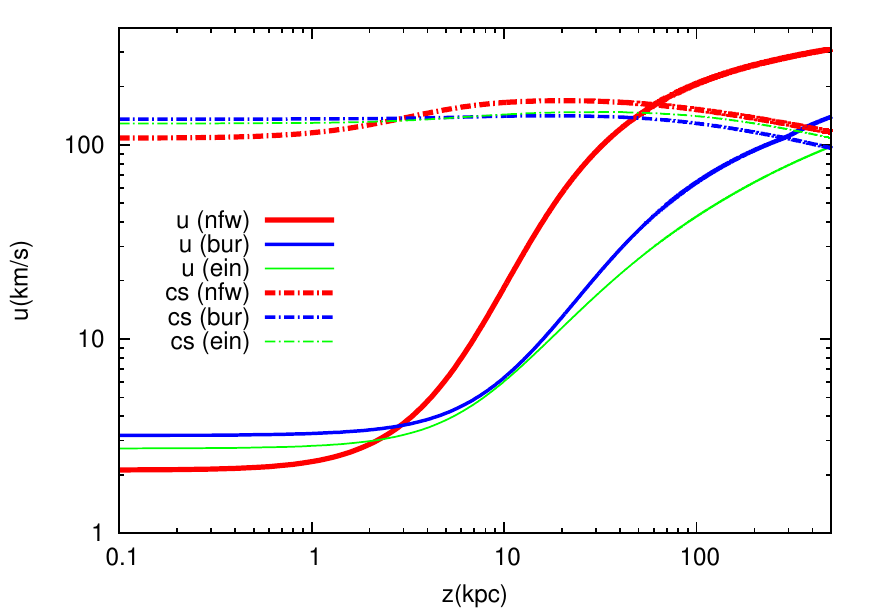}}
	}\\
	\makebox[\linewidth][c]{%
		\subfigure[Wind and Alfv\'{e}n speed.\label{fig:s-DM-u-vA}]%
		{\includegraphics[width=\linewidth]{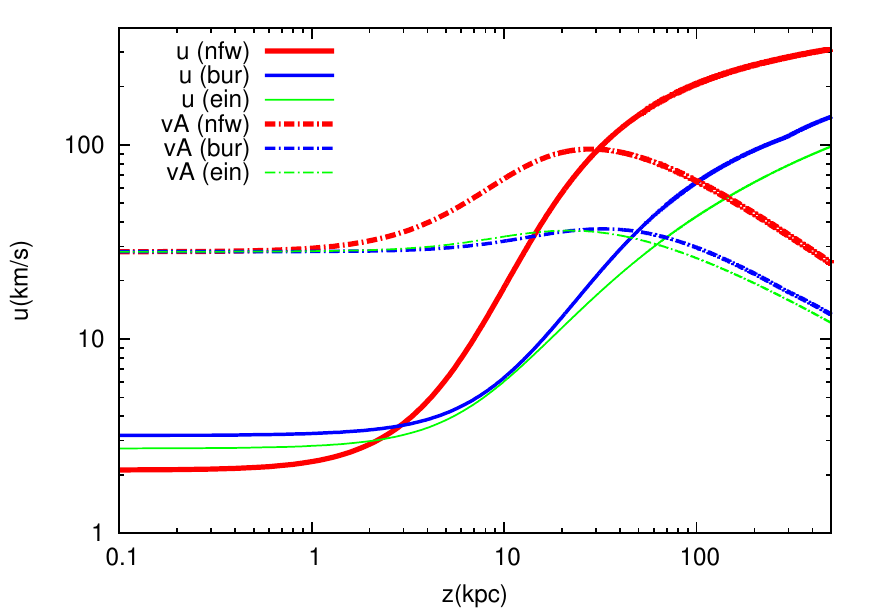}}
	}
	\caption{Dependence of the CR spectrum on the DM halo: a) wind and compound sound speed; b)  wind and Alfv\'{e}n speed.\label{fig:s-DM-A}}
\end{figure}
\begin{figure}
	
	\makebox[\linewidth][c]{%
		\subfigure[Gas density and temperature.\label{fig:s-DM-n}]%
		{\includegraphics[width=\linewidth]{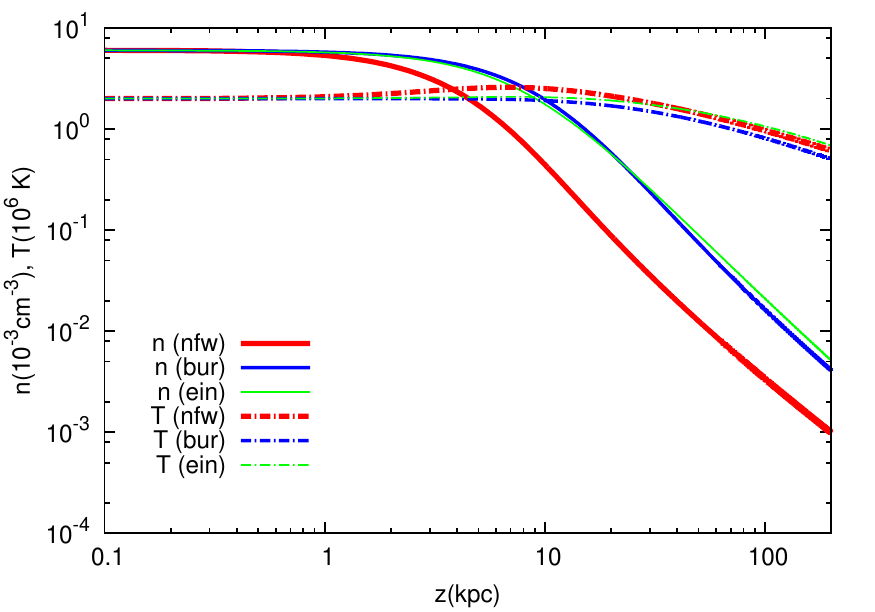}}
	}\\
	\makebox[\linewidth][c]{%
		\subfigure[Gas and CR pressure.\label{fig:s-DM-P}]%
		{\includegraphics[width=\linewidth]{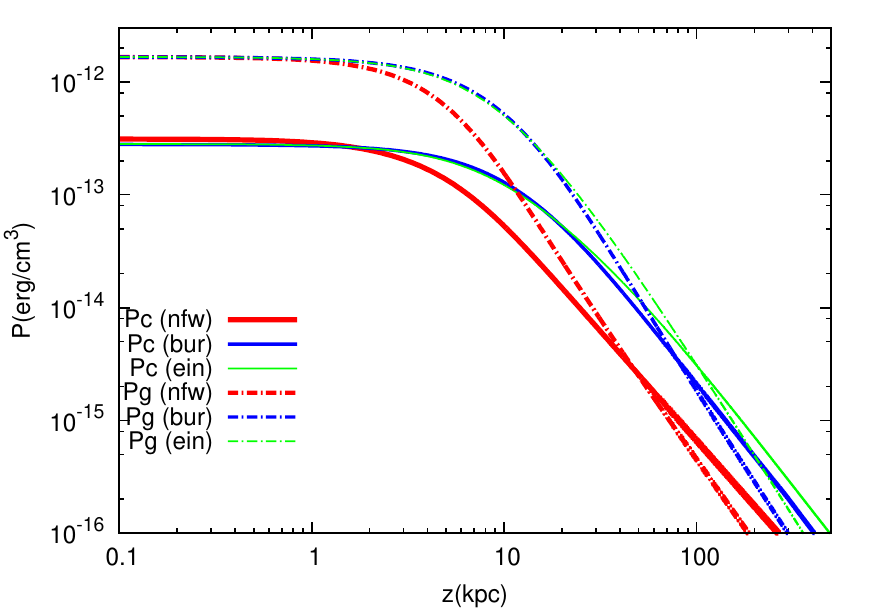}}
	}
	\caption{Dependence of the CR spectrum on the DM halo: a) gas density and temperature; b) gas and CR pressure.\label{fig:s-DM-B}}
\end{figure}
\begin{figure}
	\makebox[\linewidth][c]{%
		\subfigure[NFW-Sofue halo. \label{fig:nfw-s}]%
		{\includegraphics[width=\linewidth]{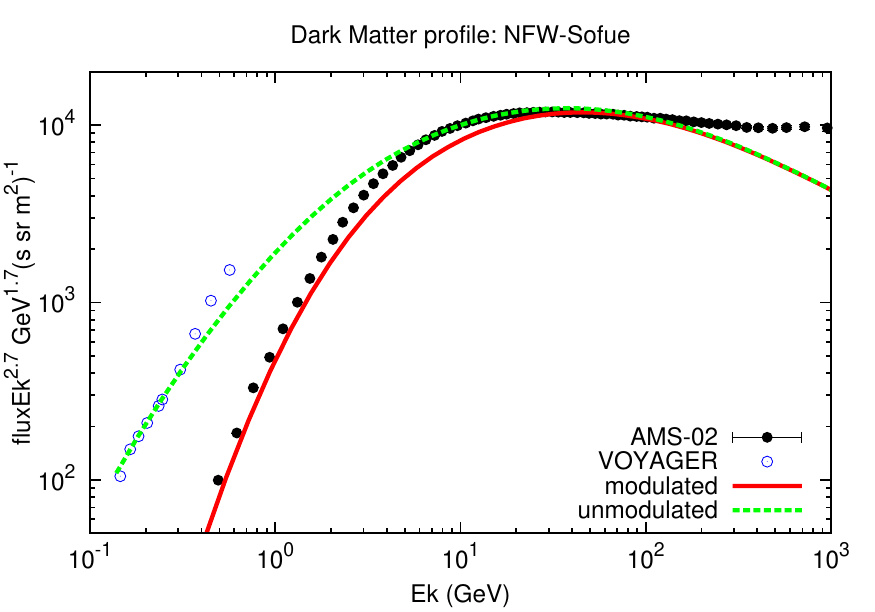}}
	}\\
	\makebox[\linewidth][c]{%
		\subfigure[Burkert halo. \label{fig:bur-s}]%
		{\includegraphics[width=\linewidth]{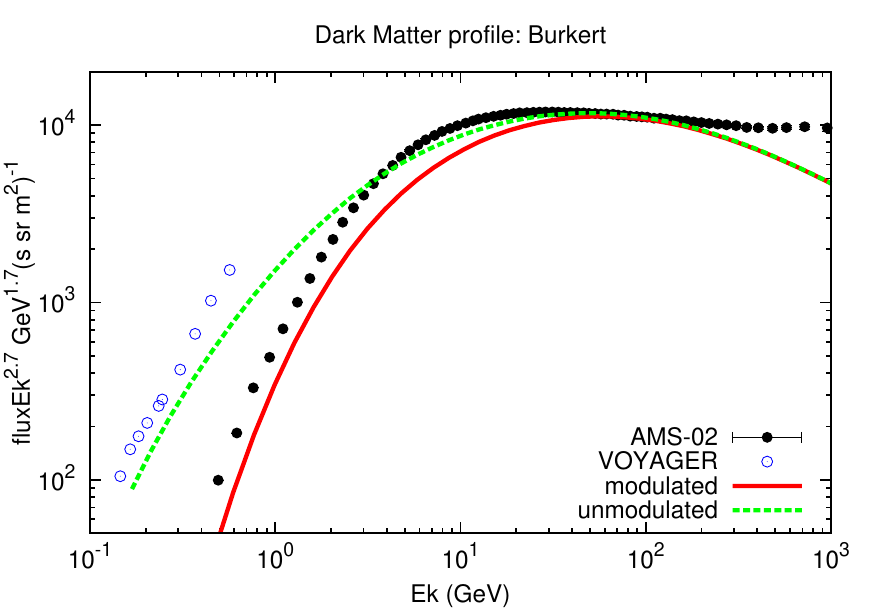}}
	}\\
	\makebox[\linewidth][c]{%
		\subfigure[Einasto halo. \label{fig:ein-s}]%
		{\includegraphics[width=\linewidth]{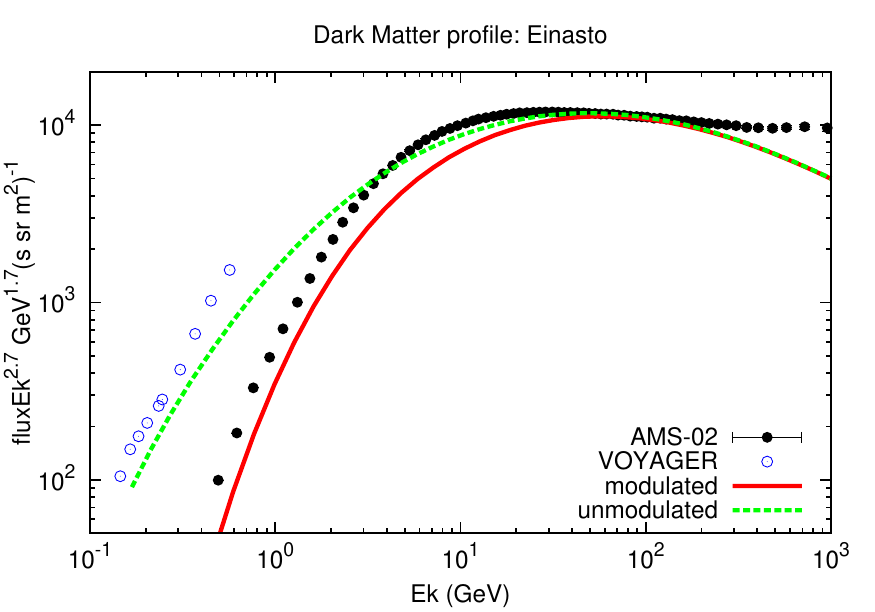}}
	}
	\caption{Dependence of the CR spectrum on the DM halo: a) CR spectrum with the NFW halo; b)  CR spectrum with the BUR halo; c) CR spectrum with the EIN halo.\label{fig:s-DM-C}}
\end{figure}

In \fig \ref{fig:s-DM-n} we show the wind density and temperature obtained with the input parameters of Model-C and for the three DM halos NFW-Sofue, BUR and EIN. The NFW-Salucci model is missing since with such halo no wind was launched. This is a typical example of how, fixed any other parameter, an increase of the gravitational pull (the NFW-Salucci has the largest force among the proposed models) could prevent the wind launching. In \fig \ref{fig:s-DM-u-cs} and \ref{fig:s-DM-u-vA} we show the corresponding wind speed, sound speed and Alfv\'{e}n velocity, while in \fig \ref{fig:s-DM-P} we show the gas and CR pressures.
For all the three DM models, the wind is launched as sub-Alfv\'{e}nic with: $(u_0 = 2\, \rm km/s,\, z_c=60 \rm kpc, \, u_f = 450 \rm km/s)$ for the NFW-Sofue profile, $(u_0 = 3\, \rm km/s,\, z_c=285 \rm kpc, \, u_f = 273 \rm km/s)$ for the BUR profile and $(u_0 = 2.7\, \rm km/s,\, z_c=590 \rm kpc, \, u_f = 253 \rm km/s)$ for the EIN profile.\\
The mass loss rate and the halo mass of the wind is  $(\dot{M} = 0.46\, \rm M_{\odot}/yr,\; M(<50 \rm kpc) \sim 1.2\times 10^9 \rm M_{\odot}, \; M(<250 \rm kpc) \sim 1.9\times 10^9 \rm M_{\odot})$ for the NFW-Sofue profile, $(\dot{M} = 0.7\, \rm M_{\odot}/yr,\; M(<50 \rm kpc) \sim 3.2\times 10^9 \rm M_{\odot}, \; M(<250 \rm kpc)\sim 6.4\times 10^9 \rm M_{\odot})$ for the BUR profile and $(\dot{M} = 0.63\, \rm M_{\odot}/yr,\; M(<50 \rm kpc) \sim 3.1\times 10^9 \rm M_{\odot}, \; M(<250 \rm kpc) \sim 8\times 10^9 \rm M_{\odot})$ for the EIN profile. Such values of the halo mass are of the same order of magnitude of the Galactic halo mass estimated by \cite{Miller-2015-0004-637X-800-1-14} $(M(<50 \rm kpc) \sim 3.8\times 10^9 \rm M_{\odot}, \; M(<250 \rm kpc) \sim 4.3	\times 10^{10} \rm M_{\odot})$. 

Notice that at larger gravitational acceleration, i.e for the NFW-Sofue profile, we get a smaller launching velocity $u_0$ and a larger terminal velocity $u_f$. Moreover, we get a smaller $z_c$. This fact can be explained as follows: in the subsonic regime the gravitational term dominates the numerator of the wind \eq \ref{eq:Hcal-wind eq}, while $c^*$ is weakly dependent on
the DM profile (see \fig \ref{fig:s-DM-A}). Thus a larger gravitational force leads to a
larger gas acceleration in the subsonic region, large enough to provide a smaller
$z_c$ despite the smaller launching velocity, as it can be seen in \fig \ref{fig:s-DM-A}.
The rapid increase of the wind velocity, which starts at smaller $z$ and with a steeper profile for the NFW-Sofue, corresponds to the rapid decrease of the gas density and of the CR and gas pressure (see \fig \ref{fig:s-DM-n} and \fig \ref{fig:s-DM-P}). In the range $z \sim 1-100$ kpc, the NFW-Sofue profile shows the largest Alfv\'{e}n speed (due to the smaller gas density), and the smallest CR pressure. However the first effect is still dominant, leading to a larger heating due to wave damping (recall that the heating term is  $\sim v_{A}\nabla P_c$), and hence to a slightly larger temperature (see \fig \ref{fig:s-DM-n}) compared to the BUR and EIN halos. Notice that the effect of wave damping is in fact to keep the gas temperature around $2 \times 10^6$ K up to $\sim 100$ kpc (this is especially true for the NFW-Sofue profile), in line with the results of \cite{Miller-2015-0004-637X-800-1-14} on the Galactic halo temperature.

In \fig \ref{fig:s-DM-C}, we show the CR spectrum for the NFW-Sofue, BUR and EIN profiles. Below $ \sim 200$ GeV, the best (qualitative) match to the observed proton spectrum is achieved with the NFW-Sofue model. The larger wind launching velocity of the BUR and EIN models leads to a low energy spectrum harder than the observed one.

\section{Conclusions}
In this paper we studied the possible presence of CR-driven Galactic winds launched at the Sun location and the dependence of their properties and of the related CR spectrum on the conditions of the Galactic environment in the vicinity of the Sun. We used the semi-analytical calculation developed in \citetalias{Recchia-2016-08082016}, which allows to compute at the same time both the hydrodynamical structure of the wind, the CR distribution function and CR self-generated  diffusion coefficient. The CR transport is considered to be due to the diffusion off Alfv\'{e}n waves generated by CRs through streaming instability, saturated through NLLD, and to the advection with such waves and with the wind. The wind launching and the CR spectrum depend on the properties of the ISM (gas density and temperature, Galactic magnetic field), on the CR injection, on the flow geometry and on the Galactic gravitational potential. All these quantities depend
on the position in the Galaxy and are constrained by observations.\\

We found that, in agreement with previous hydrodynamic  calculations of CR-driven winds, the combined action of the thermal and CR pressures can drive winds for a variety of environmental conditions. In particular, it is possible to launch winds with input parameters compatible with observations of the ISM in the vicinity of the Sun. We also found that in many cases, winds launched with environmental parameters compatible with observations lead to a CR spectrum which is not in agreement with the observed one. As pointed out in \citetalias{Recchia-2016-08082016}, this is mainly due to the fact that CR advection in such winds is strong, leading to spectra at low energies (below $\sim 200$ GeV) which are harder than the observed spectrum. At high energy (above $\sim 200$ GeV), instead, the wind expansion, together with the steep energy dependence of the self-generated diffusion coefficient, leads to spectra which are steeper than the observed spectrum. However, the high energy spectrum may be affected by other factors, such as the presence of turbulence non generated by CRs (\cite{Aloisio-2015p3650}; \citetalias{Recchia-2016-08082016}).\\
As for the low energy part of the spectrum, we found that it is possible to find specific cases in which the wind launching parameters and the resulting wind properties are compatible with observations, in particular with the Galactic halo properties deduced by \cite{Miller-2015-0004-637X-800-1-14}, and the resulting CR spectrum is also in agreement with observations. In this analysis we also found that an important role is played by the choice of the dark matter halo model. In fact, we showed that, for given values of all parameter, the DM gravitational potential can determine whether the wind is launched and the quality of the agreement with the observed CR spectrum.\\
This analysis highlights that the low energy part of the CR spectrum provides a strong constraint for  wind models and that an accurate modeling of the Galactic environment and of the Galactic gravitational potential (in particular of the DM halo) are of crucial importance in the understanding of Galactic winds.\\

The study presented here can be extended to any other locations in the Galaxy different from the Sun's position and can be used to analyze the possible presence of winds and the related CR spectrum in the whole Galaxy. Such investigation would allow us to calculate the rate of mass loss and the mass of the baryonic halo induced by the wind (both are crucial ingredients in models of galaxy formation and evolution), as well as the expected intensity in emission and absorption lines from the halo (for instance in the X-ray band) and the CR density gradient in the Galaxy (which can be tracked by the $\gamma-$ray emission resulting from  CR interactions with the background plasma).     
\section*{Acknowledgements}

The authors are glad to acknowledge a continuous ongoing
collaboration with the rest of the Arcetri high energy Group. G.M. want also to acknowledge the hospitality of ISSI for Teamwork 351 on "The origin and composition of Galactic cosmic rays". 




\bibliographystyle{mnras}
\bibliography{biblio} 






\bsp	
\label{lastpage}
\end{document}